%% file: main.tex
\documentclass[11pt]{article}

\input{preamble}

\title{Antitrust on Aisle Five: \\ How Well Do Divestiture Remedies Work?\footnote{We thank Paul Frangie and Joshua Smith for their help in data collection, and Angelina Tomseth for her excellent research assistance. We also thank Marc Remer, Ted Rosenbaum, Dave Schmidt, and David Shaw for their comments on this paper. This research was conducted with restricted access to Bureau of Labor Statistics (BLS)
data. The views expressed in this paper are those of the authors. They do not necessarily represent those of the Bureau of Labor Statistics (BLS), the Federal Trade Commission (FTC), or any of the FTC Commissioners. This paper provides a summary of research results. The information is being released for statistical purposes, to inform interested parties, and to encourage discussion of work in progress. The paper does not represent an existing or a forthcoming new official BLS statistical data product or production series. All results have been reviewed to ensure no confidential information is disclosed. Analysis with BLS data was performed by Dominic Smith. Analysis with FTC data was performed by the non-BLS authors. References to specific companies are based on public information and do not imply the company is in any of the confidential data sources.} 
}

    \author{Xiao Dong\thanks{Federal Trade Commission. Email: \texttt{xdong@ftc.gov}; \texttt{devesh.raval@gmail.com}; \texttt{bwendling@ftc.gov}. } \and Paul Koh\thanks{Yonsei University. Email: \texttt{paulkoh9@gmail.com}.} \and Devesh Raval\footnotemark[2] \and Dominic Smith\thanks{Case Western Reserve University. Email: \texttt{das61@case.edu}. } \and Brett Wendling\footnotemark[2]}
\date{
April 8, 2026
}

\begin{document}

\maketitle
\input{abstract} \clearpage
\input{Introduction}
\input{Background}

\input{Results}
\input{Mechanism}
\input{Conclusion}

\singlespacing
\bibliographystyle{econ}
\bibliography{references}
\onehalfspacing

\clearpage
\appendix
\part*{Appendix}
\input{App.controls}

\clearpage
\singlespacing

\end{document}

%% file: preamble.tex
\usepackage{graphicx} %
\usepackage[utf8]{inputenc}
\usepackage{eurosym,geometry,graphicx,color,setspace,sectsty,comment,footmisc,natbib,pdflscape,array,hyperref}
\usepackage[normalem]{ulem}
\usepackage{caption}
\usepackage{subcaption} %
\usepackage{amssymb,amsfonts,amsmath}
\usepackage{amsthm}
\usepackage{natbib}
\usepackage{booktabs}
\usepackage{threeparttable}
\usepackage{tikzsymbols} %
\usepackage{tikz}
\usetikzlibrary{patterns,arrows}
\usepackage{apxproof}
\usepackage{xurl}
\usepackage{clipboard}
\usepackage{xcolor} %
\usepackage{todonotes}
\setuptodonotes{inline}
\usepackage[]{mdframed}

\newtheorem{assumption}{Assumption}
\newtheorem{lemma}{Lemma}
\newtheorem*{lemma*}{Lemma}

\newtheorem*{theorem*}{Theorem}
\newtheorem{corollary}{Corollary}
\newtheorem{proposition}{Proposition}
\newtheorem{claim}{Claim}

\definecolor{webpurple}{rgb}{0.7,0,0.7}

\newcommand{\commentout}[1]{}
\newcommand{\propref}[1]{\hyperref[#1]{Proposition~\ref*{#1}}}
\newcommand{\lemmaref}[1]{\hyperref[#1]{Lemma~\ref*{#1}}}
\newcommand{\assumptionref}[1]{\hyperref[#1]{Assumption~\ref*{#1}}}
\newcommand{\corollaryref}[1]{\hyperref[#1]{Corollary~\ref*{#1}}}
\newcommand{\claimref}[1]{\hyperref[#1]{Claim~\ref*{#1}}}
\newcommand{\appref}[1]{\hyperref[#1]{Appendix~\ref*{#1}}}
\newcommand{\wappref}[1]{\hyperref[#1]{Web Appendix~\ref*{#1}}}
\newcommand{\secref}[1]{\hyperref[#1]{Section~\ref*{#1}}}
\newcommand{\figref}[1]{\hyperref[#1]{Figure~\ref*{#1}}}
\newcommand{\tabref}[1]{\hyperref[#1]{Table~\ref*{#1}}}
\newcommand{\defref}[1]{\hyperref[#1]{Definition~\ref*{#1}}}

\renewcommand{\eqref}[1]{\hyperref[#1]{\textcolor{black}{(}\ref*{#1}\textcolor{black}{)}}}
\newcommand{\neqref}[1]{\hyperref[#1]{equation~\ref*{#1}}}
\newcommand{\Eqref}[1]{\hyperref[#1]{Equation~\textcolor{black}{(}\ref*{#1}\textcolor{black}{)}}}
\newcommand{\numref}[1]{\hyperref[#1]{\textcolor{black}{(}\ref*{#1}\textcolor{black}{)}}}

\theoremstyle{definition} %
\newtheorem{definition}{Definition}

\newtheorem{example}{Example}
\newtheorem*{example*}{Example}

\AtBeginEnvironment{example}{%
  \pushQED{\qed}%
}
\AtEndEnvironment{example}{\popQED\endexample}

\AtBeginEnvironment{remark}{%
  \pushQED{\qed}%
}
\AtEndEnvironment{remark}{\popQED\endexample}

\definecolor{low_gray}{gray}{0.85}
\definecolor{low_gray2}{gray}{0.75}
\definecolor{low_gray0}{gray}{0.75}
\definecolor{mid_gray}{gray}{0.65}
\definecolor{mid_gray2}{gray}{0.55}
\definecolor{mid_gray3}{gray}{0.45}
\definecolor{seagreen}{cmyk}{0.669,0.000,0.374,0.455}
\definecolor{webgreen}{rgb}{0,.5,0}
\definecolor{webbrown}{rgb}{.6,0,0}
\definecolor{webyellow}{rgb}{0.98,0.92,0.73}

\hypersetup{pdfpagemode=, colorlinks=true,
anchorcolor= webbrown, citecolor= webbrown,
filecolor= webbrown, linkcolor= webbrown, menucolor= webbrown,
urlcolor= webbrown, citebordercolor= 1 0 0, menubordercolor=1 0
0, urlbordercolor=1 0 0, runbordercolor=1
0 0}
\usepackage{xurl} %

%% file: abstract.tex
\begin{abstract}
Antitrust authorities frequently rely on structural divestitures to address competitive concerns raised by mergers. Using census-level establishment data and proprietary transaction records from the U.S. grocery sector, we provide systematic evidence on the long-run effects of such remedies. Divested stores experience an average 31 percent decline in employment over five years, driven by elevated exit rates and persistent contraction among surviving establishments. Sales similarly decline. Transaction-level evidence indicates that divested assets are systematically weaker and are often transferred to lower-capability buyers. These findings suggest that structural remedies may be less effective when the implementation of divestitures allows merging parties substantial discretion over the assets and buyers involved. %


\end{abstract}

%% file: Introduction.tex
 \section{Introduction}
 
\Copy{first_divest_year}{1990}
\Copy{last_divest_year}{2015}
\Copy{num_divest_estabs}{700}
\Copy{num_divestitures}{25}
\Copy{num_sold}{10}
\Copy{num_not_sold}{15}
\Copy{base_year}{20}

When reviewing a merger, antitrust agencies frequently face a choice between blocking the transaction and approving it subject to remedies. The most common structural remedy is divestiture: requiring the merging firms to sell assets to independent buyers. This approach assumes that transferring assets to a rival can replicate the competitive constraint that would otherwise be eliminated by the merger.

Whether divestitures in fact restore competition is theoretically ambiguous and empirically unresolved. Because merging firms continue to compete against the divested assets, they may have incentives to structure remedies in ways that weaken the competitive threat posed by the divestiture buyer. For retail divestitures, for example, firms may divest stores with weaker fundamentals or sell to buyers with limited operational capabilities. If buyer quality and asset composition affect post-divestiture performance, the competitive constraint exerted by divested stores may differ from that of the pre-merger firms.

The U.S. supermarket industry provides a useful setting to study the effectiveness of merger remedies. Competition in grocery retail is highly local, so mergers frequently require divestitures of overlapping stores in narrowly defined geographic markets. Since 1990, the Federal Trade Commission has ordered divestitures of more than 600 supermarkets across over 25 merger investigations, creating repeated opportunities to evaluate remedy performance. At the same time, the industry is economically important, employing nearly 3 million workers and generating more than \$800 billion in annual sales in 2022—approximately 12 percent of all retail trade—making the consequences of merger remedies economically meaningful.

The relevance of these questions has increased as grocery retail has become more concentrated over time, with sales shifting toward national chains partly through mergers of regional operators \citep{smith2025evolution}. For example, the proposed Kroger–Albertsons merger would have continued this trend by creating the largest supermarket chain in the United States. Litigation over the proposed transaction focused in part on whether the proposed divestiture buyer would be able to operate the divested stores as effective competitors, highlighting the importance of understanding whether stores remain viable after divestiture \citep{chesnes2025economics}.

This paper provides a systematic evaluation of the longer-run effectiveness of divestiture remedies. Using establishment-level data, we study the post-divestiture performance of supermarkets in terms of survival, employment, earnings, sales, and profit margins relative to comparable control stores. We then examine the mechanisms underlying these outcomes, focusing on asset selection, buyer capability, and transition dynamics.

Using restricted-access, administrative establishment-level data from the U.S. Bureau of Labor Statistics, we find that divested stores experience substantial performance declines following transfer. Relative to comparable control stores, closure rates are on average 13 percentage points higher over the first year and 17 percentage points higher over the first five years. Conditional on survival, employment is on average 25 percent lower over the first year and 17 percent lower over the first five years. Accounting for both store exit and employment changes among surviving stores, total employment at divested stores declines by 31–33 percent on average over one- to five-year horizons. Average worker earnings at divested stores also decline by approximately 7 percent on average over five years.

The multi-decade span of our data allows us to examine whether divestiture performance has improved over time. Stores divested in the later period exhibit substantially higher exit rates in the first year after divestiture than stores divested earlier. However, exit rates among stores divested in the earlier period continue to rise over time, so that survival differences between the two periods narrow by five years after divestiture.

Conditional on survival, employment dynamics also differ across periods. In the earlier period, employment among surviving stores remains substantially below baseline even after five years, whereas in the later period it converges to levels comparable to controls. Combining exit and conditional employment effects, the average employment decline over five years is 28 percent in the early period and 35 percent in the later period, suggesting similar overall outcomes of supermarket divestitures over time.

We complement these findings with conclusions drawn from analysis of proprietary store-level financial data from supermarket chains. In two merger case studies, sales at divested stores decline by 20 to 30 percent following divestiture. In one case, profit margins fall by nearly 80 percent. These patterns mirror the employment evidence and indicate that divested stores often operate at substantially reduced scale and profitability following divestiture.

We next investigate mechanisms underlying these patterns. Both asset selection and buyer capability appear central to divestiture performance. First, divested stores are systematically weaker prior to transfer. In transaction-level data, divested stores exhibit substantially lower pre-divestiture sales and profit margins than nearby retained stores, consistent with selective divestiture of lower-performing assets. Second, buyer quality matters. In two transactions, the acquiring firms operate stores with lower sales than the merging parties prior to divestiture. Moreover, within a single divestiture involving multiple buyers, stores transferred to smaller buyers experience persistent sales declines of 33-36 percent even a decade after transfer, whereas stores transferred to larger buyers exhibit increases in sales relative to retained stores.

By contrast, direct transition frictions appear limited. Remodeling costs are modest relative to annual sales, and ownership changes unrelated to divestitures have negligible effects on store performance in most cases we examine. Buyer expectations and forecasting sophistication vary substantially across transactions. In one transaction, a buyer substantially overestimated post-transfer sales, whereas a more experienced firm explicitly projected several years of below–steady-state performance, reflecting a more sophisticated assessment of the operational adjustment required following asset transfer. Overall, the evidence points to persistent differences in asset quality and buyer capability—rather than short-run transition costs—as the primary drivers of post-divestiture underperformance.

A substantial empirical literature evaluates the price and output effects of consummated mergers across industries. By contrast, relatively little work studies whether divestitures—the primary remedy used to address competitive harm—preserve the competitive capacity lost in mergers. On the theory side, \citet{loertscher2024mergers} and \citet{nocke2025optimal} analyze the optimal design of divestiture remedies. On the empirical side, existing work primarily examines individual divestiture events in specific industries and has generally found that divestitures successfully maintained competition in affected markets.\footnote{For example, see airlines \citep*{zhang2017effects}, beer \citep*{friberg2015divestiture, wang2023divestiture}, casinos \citep*{osinski2021evaluating}, coffee \citep*{delaprez2024merger}, consumer packaged goods \citep*{delaprez2024unveiling}, electricity \citep*{brown2023evaluating}, pharmaceuticals \citep*{chen2022competitive, tenn2011success}, retail gasoline \citep*{soetevent2014auctions, lagos2018effectiveness}, and tobacco \citep*{nathan2025can}.} While these studies provide valuable case-specific evidence, there is limited systematic evidence on the long-run performance of divested assets and on the mechanisms determining whether they remain viable competitors.

One exception is \citet{chen2022competitive}, who study FTC-mandated divestitures of generic drugs resulting from mergers between 2005 and 2016. They find that affected markets have fewer competitors two to four years after divestiture relative to comparable markets, driven by both increased firm exit and reduced entry. They also estimate price increases of 2–6 percent, although these effects are not statistically significant.

Two FTC retrospective reports \citep{ftc1999divestiture, ftc2017merger} evaluate divestiture outcomes using case-study methodologies covering transactions from 1990–1994 and 2006–2012, respectively. The earlier study concludes that 75 percent of divestitures were successful, while the later study finds that 69 percent were successful and an additional 14 percent were ``qualified successes."\footnote{\citet{ftc1999divestiture} classify a divestiture as successful if the buyer continued operating in the relevant market. \citet{ftc2017merger} define success as competition remaining at or returning to pre-merger levels within two to three years, and ``qualified success" if competition returned after a longer period.} Although the definitions of success differ across reports, both conclude that divestiture remedies were generally effective while recommending refinements to increase the likelihood of success. In contrast to these case-based evaluations, we provide a systematic, establishment-level analysis of divestiture outcomes using a consistent empirical framework and longer post-divestiture horizons.

Recent work has begun to examine supermarket divestitures directly. \citet*{hosken2024does} analyze two major supermarket divestitures using online Yelp reviews and find that negative reviews increase following divestiture, particularly with respect to pricing complaints. These findings are consistent with a decline in competitive pressure, but they focus on consumer perceptions rather than the underlying performance and viability of divested stores. \citet{argentesi2021effect} concludes that a divestiture of supermarkets in the Netherlands was successful. \citet{koh2025merger} shows that the FTC's divestiture remedy in the Albertsons--Safeway merger moderately decreased the upward pricing pressures.

Our findings also connect to the broader literature on asset specificity and redeployability of productive assets \citep*{williamson1975markets, williamson1985economic, williamson1986economic}. Theoretical and empirical work emphasizes that assets may be highly firm-specific, limiting their value outside the originating firm \citep{kermani2023asset, kim2018asset}. Our setting provides a novel test of this idea in the context of regulatory asset transfers. Because grocery markets are local and buyer pools are limited, divested stores may be particularly sensitive to buyer capability and asset complementarities. The finding that divested stores often underperform following transfer is consistent with the presence of firm-specific assets that are not easily redeployed across firms.

The rest of the paper is organized as follows. \secref{section:background} provides institutional background on divestitures and grocery merger enforcement. \secref{section:results} studies the effect of divestiture on store survival and performance. \secref{section:mechanism} explores the economic mechanisms driving the loss of divested stores' ability to compete effectively. \secref{section:conclusion} concludes. 

%% file: Background.tex
\section{Background\label{section:background}}

\subsection{Divestitures as a Merger Remedy}

Mergers in the United States are evaluated under Section 7 of the Clayton Act, which prohibits transactions whose effect ``may be substantially to lessen competition." In practice, antitrust agencies assess whether a proposed merger is likely to increase market power in affected markets. When a merger generates efficiencies but raises competitive concerns in a subset of markets, agencies face a tradeoff between blocking the transaction entirely and allowing efficiencies to be realized \citep{williamson1968economies}.

One way to address this tradeoff is through merger remedies. Instead of prohibiting the transaction outright, agencies may require targeted modifications that address the sources of competitive harm while allowing the remainder of the merger to proceed. Such remedies allow agencies to address competitive concerns in specific markets while preserving efficiencies elsewhere. Remedies may also reduce litigation risk and enforcement costs. If a court declines to block the merger, the transaction proceeds as proposed, leaving the anticipated competitive harm unaddressed, while litigating merger challenges entails substantial direct costs such as expert witness fees and opportunity costs when agency staff devote months preparing for trial.

Divestiture remedies operationalize this approach. Rather than prohibiting the merger, agencies permit the transaction to proceed conditional on the sale of assets in affected markets to an independent buyer. The objective is to preserve competition where harm is anticipated while allowing the merged firm to operate elsewhere. In industries such as supermarkets, where competition is local and geographic overlaps are limited, divestitures have been a commonly utilized remedy.

From an economic perspective, divestitures pose a delegation problem. Although regulators specify the objective of preserving competition, merging firms typically propose the asset package and negotiate the sale to a buyer \citep{FTC2012negotiating}.\footnote{The complaint filed by Albertsons following the collapse of the proposed Kroger–Albertsons merger provides a detailed description of the process Kroger undertook when proposing an asset package and buyer to the FTC. See  \citet{albertsons_kroger_complaint_2024} for additional details.} Because the merged firm will compete against the divested assets post-transaction, its incentives may not align with the regulator’s objective. 

Agencies must therefore evaluate whether the proposed divestiture will preserve competition and decide whether to accept the settlement. A recent statement by FTC Chairman Ferguson summarizes the relevant criteria: the divested assets must constitute a viable standalone business capable of competing vigorously with the merged firm, and the buyer must possess sufficient resources and experience.\footnote{Chairman Ferguson said: 
\begin{quote}
    Nor should the Commission ordinarily accept a structural remedy unless it involves the
    sale of a standalone or discrete business, or something very close to it, along with all
    tangible and intangible assets necessary (1) to make that line of business viable, (2)
    to give the divestiture buyer the incentive and ability to compete vigorously against
    the merged firm, and (3) to eliminate to the extent possible any ongoing entanglements
    between the divested business and the merged firm. The Commission must also be
    confident that the divestiture buyer has the resources and experience necessary to make
    that standalone business competitive in the market.
\end{quote} See \url{https://www.ftc.gov/system/files/ftc_gov/pdf/synopsys-ansys-ferguson-statement-joined-byholyoak-meador.pdf}. Commissioner Mark Meador lays out the principles he considers when evaluating divestitures in \url{https://www.ftc.gov/system/files/ftc_gov/pdf/mark-meador-statement-act-giant-eagle.pdf}.
}

Divestiture effectiveness thus depends on the interaction between asset quality and buyer capability. First, if stores differ in underlying quality, such as their location, productivity, or customer loyalty, the merging firm may have incentives to divest assets with weaker fundamentals. Second, the competitive effects of a divestiture depend on the buyer’s operational expertise, brand strength, scale, and ability to integrate the assets into existing systems. If assets are partially firm-specific or rely on complementary capabilities such as supply chains or private-label programs, their value may decline when transferred to a less capable buyer. When assets are imperfectly redeployable, buyer pools are limited, or the seller influences the matching process, the resulting allocation may not replicate the competitive constraint of the pre-merger firms.

\subsection{Divestiture Remedies in the Supermarket Industry}

These considerations are particularly salient in the supermarket industry. Competition is local, stores are heterogeneous in quality, and performance depends on access to distribution networks, private-label programs, managerial expertise, and brand reputation. The pool of potential buyers in a given geographic area is often limited to existing regional chains or smaller operators. As a result, divestitures in grocery markets provide a setting in which the incentives surrounding asset selection and buyer assignment are likely to be economically meaningful.%

Supermarket mergers have rarely been challenged. Since 1990, the FTC has litigated only two supermarket mergers. In most cases, agencies have relied on negotiated structural remedies rather than outright prohibition. The scale and complexity of divestiture packages have increased alongside merger size. This pattern underscores the central role of divestitures in grocery merger enforcement and motivates a systematic evaluation of their effectiveness.

%% file: Results.tex
\section{Effects of Divestiture \label{section:results}}

\subsection{Divested Stores\label{sub:construction}}

To evaluate the post-divestiture performance of grocery stores, we link our list of divested establishments to the Bureau of Labor Statistics’ Quarterly Census of Employment and Wages Longitudinal Database (QCEW LDB), which provides quarterly wage and employment information for the universe of U.S. establishments \citep*{pivetz2001,sadeghi2016}. The QCEW LDB includes a persistent establishment identifier that allows us to track stores over time, as well as Employer Identification Numbers (EINs) that permit us to identify changes in enterprise ownership following divestitures.\footnote{Any reference to a company in another part of this paper does not imply that it is included in the analytical sample for this section. This section has been reviewed to ensure the confidentiality of all companies.}

We identify divested establishments by matching store addresses in our divestiture database to establishments in the QCEW LDB and verifying that the associated firm name corresponds to either the divesting or acquiring firm around the time of the transaction. Using this approach, we match 448 of 601 divested stores (75\%) to establishments in the QCEW LDB with positive employment four quarters prior to the divestiture.\footnote{Unmatched stores may arise from errors in our hand-constructed divestiture list, cases in which a divested store exited more than four quarters prior to the expected divestiture date, or instances in which the establishment is not classified as a grocery store in the QCEW LDB. We individually investigated each unmatched store to identify potential matches in the QCEW LDB.} Of the establishments we match, more than 95 percent exhibit a change in EIN between one quarter before and one quarter after the divestiture, which helps confirm that we have identified the correct establishment.

Our unit of analysis is the physical store location.\footnote{Specifically, we use the QCEW LDB-provided latitude and longitude coordinates rounded to four decimal places, corresponding to approximately 10–15 meters of spatial precision. Given the physical size of grocery stores and typical lot spacing, this level of precision is sufficient to distinguish distinct retail establishments, including in dense urban areas, in nearly all cases. For the small number of addresses that we cannot geocode we use the centroid of the zip code of the address.} Because establishment identifiers sometimes change around the time of divestiture, focusing on physical locations avoids conflating administrative identifier changes with true store closures.

\subsection{Control Stores\label{sub:control}}

We compare divested stores to observationally similar non-divested control stores to estimate the effect of divestiture on store performance. Control stores are selected using restrictions designed to approximate parallel trends in the absence of divestiture.

First, control stores must be located outside the three-digit ZIP codes of divested stores to limit contamination from local competitive effects. Second, controls must belong to a multi-store grocery chain, be within eight quarters of the divested store’s age, and have baseline employment within 20 percent of the divested store’s employment.\footnote{We measure employment differences using the change formula in equation \eqref{eqn:dhs}.} 

Third, to prevent control stores from being affected by merger activity, we divide divestitures into those occurring before 2008 and those occurring after 2008. Within each period, we exclude potential control stores whose EIN belongs to any firms involved in the mergers or acquisitions associated with the divested stores in that time period. In addition, control stores must not share an EIN with any divested store.

Given these restrictions, we select the ten geographically closest stores as controls for each divested store. These restrictions align treated and control stores along key observable dimensions, such as scale, organizational form, life-cycle stage, and location, while avoiding contamination from local competitive effects or merger activity affecting the treated stores.\footnote{We examine both more and less restrictive control groups in \appref{app:controls} and find that the choice of control group does not meaningfully affect our results.}

\subsection{Descriptive Evidence}

\tabref{tab:treatment_control} compares observable characteristics of divested and control stores in the period four quarters prior to divestiture. The two groups have very similar employment levels, although control stores have 9 percent lower average payroll. These differences are modest relative to the dispersion within each group, suggesting that the control stores provide a reasonable benchmark for evaluating post-divestiture outcomes.

\begin{table}[htbp!]
  \centering
  \small
  \begin{threeparttable}
    \caption{Pre-Divestiture Summary Statistics by Treatment Group}
    \label{tab:treatment_control}
    \begin{tabular}{lccccccc}
      \toprule
      & Establishments & \multicolumn{2}{c}{Employment} & \multicolumn{2}{c}{Payroll} & \multicolumn{2}{c}{Average Earnings} \\ \cmidrule(lr){2-2} \cmidrule(lr){3-4} \cmidrule(lr){5-6} \cmidrule(lr){7-8}
      Group & Count & Avg & SD & Avg & SD & Avg & SD \\
      \midrule
      Control   & 1296 & 76.95 & 27.74 & 662{,}300 & 302{,}100 & 8{,}670 & 2{,}899 \\
      Treatment &  448 & 77.05 & 27.44 & 728{,}400 & 319{,}500 & 9{,}474 & 2{,}825\\
      \bottomrule
    \end{tabular}
    \begin{tablenotes}
      \footnotesize
      \item \emph{Notes}: This table reports the number of establishments and the mean and standard deviation of employment, payroll, and average earnings, separately for treated (divested) and control stores four quarters before the divestiture. All estimates are based on data from the BLS QCEW LDB.
    \end{tablenotes}
  \end{threeparttable}
\end{table}

\begin{figure}[htbp!]
    \centering
    \begin{minipage}{1\linewidth}
    \includegraphics[width=1\linewidth]{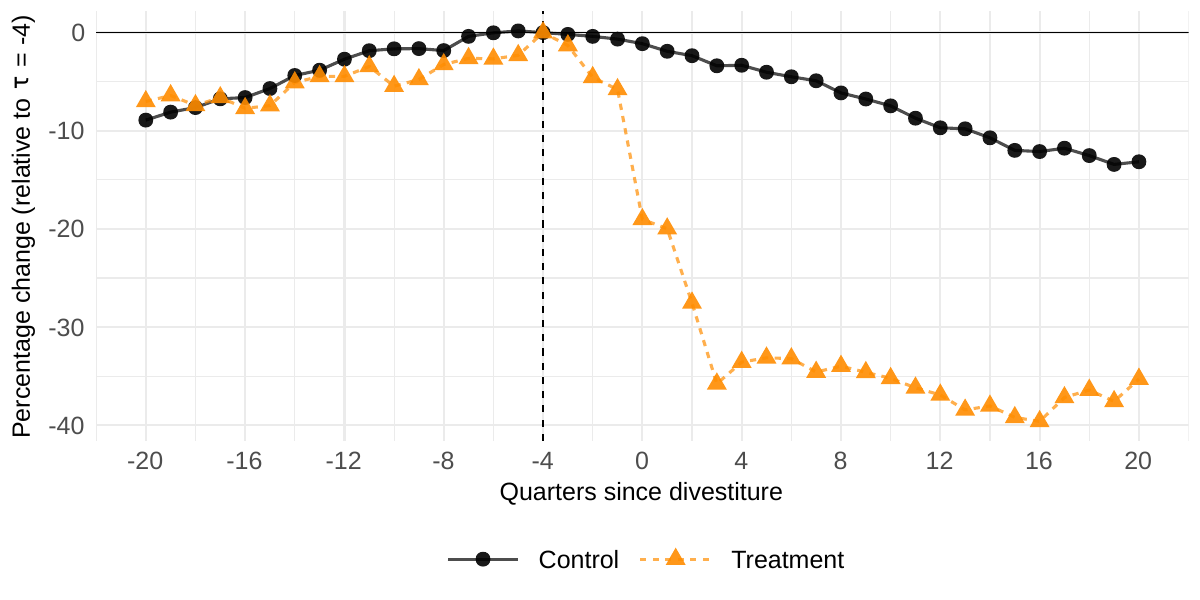}
    \caption{Change in Total Store Employment Relative to Pre-Divestiture Baseline ($\tau = -4$)} \label{fig:employment}
    \begin{flushleft}
        \footnotesize
        \textit{Notes:} All estimates are based on data from the BLS QCEW LDB. The figure shows the percentage change in sample employment for the treated and control groups relative to four quarters preceding the divestiture date.
    \end{flushleft}
    \end{minipage}
\end{figure}

We next examine how overall employment evolves for treated and control stores following divestiture. \figref{fig:employment} plots total employment by event time for divested (treated) locations (dashed orange line) and control locations (solid black line). Each series is normalized to zero at the pre-divestiture baseline ($\tau = -4$), and values are expressed as percentage deviations from that baseline. By construction, the two series coincide at $\tau = -4$, and employment trends are similar across groups prior to divestiture.

Following divestiture, employment paths diverge sharply between treated and control stores. One year after divestiture, employment at divested stores declines by roughly one-third, compared with only a 3 percent decline at control stores. The gap persists over time: five years after divestiture, employment at divested stores remains 34 percent below baseline, compared with a 13 percent decline at control stores. We next decompose this divergence into the contributions of store exit and employment changes among surviving establishments using a stacked difference-in-differences framework.

\subsection{Identification Strategy}

Our identification strategy exploits variation in the timing of divestitures across store locations. We compare the evolution of outcomes at divested stores to observationally similar control stores before and after divestiture using a stacked difference-in-differences framework following \citet*{wing2024stacked}. In this approach, each divestiture event defines a cohort (sub-experiment). We stack balanced event-time panels across cohorts and weight treated and control observations to maintain compositional balance across treatment cohorts.

Identification relies on two standard assumptions. First, a modified no-anticipation assumption requires that divestiture does not affect store outcomes prior to the baseline period used in the analysis. Second, the parallel trends assumption requires that, absent divestiture, outcomes at treated and control stores would have evolved similarly over time. Because firms may anticipate which stores will ultimately be divested and begin adjusting operations prior to transfer, we normalize outcomes to four quarters before divestiture and define event time relative to the transfer date. This specification allows for potential adjustments in the quarters immediately preceding divestiture while assessing the identifying assumptions using earlier pre-divestiture event-time coefficients.

Let $g$ index each sub-experiment cohort defined by a divestiture event. We set the event window to $\kappa = 20$ quarters, so each cohort is observed over a balanced panel of length $2\kappa + 1$, indexed by event time $\tau \in \{-\kappa, \ldots, \kappa\}$. We estimate the following stacked event-study specification:
\begin{equation}\label{eq:event_study}
Y_{ig\tau} = \alpha_0 + \alpha_1 T_{ig} + \sum_{k \neq -4} \Big( \gamma_k \mathbf{1}\{\tau = k\} + \beta_k T_{ig} \cdot \mathbf{1}\{\tau = k\} \Big) + u_{ig\tau},
\end{equation}
where $Y_{ig\tau}$ is the outcome for location $i$ in cohort $g$ at event time $\tau$, and $T_{ig}$ indicates whether location $i$ is divested in cohort $g$. The coefficients $\beta_k$ trace the dynamic treatment effects relative to the omitted baseline period $\tau = -4$. We cluster standard errors at the three-digit ZIP-code level to account for spatial correlation in outcomes. We estimate \eqref{eq:event_study} using weighted least squares, with weights constructed to balance treatment-group and control-group trends within each cohort. 

To summarize magnitudes in a compact format, we aggregate the event-study coefficients $\beta_k$ for a given horizon. For each horizon $h \in \{4,8,20\}$ quarters, we compute the average treatment effect over the post-divestiture window $k \in [1,h]$ as
\begin{equation}\label{eq:lin_com}
\frac{1}{h}\sum_{k=1}^h \beta_k,
\end{equation}
where the effect remains relative to the omitted baseline period $\tau = -4$. Averaging across post-divestiture periods reduces noise in individual event-time coefficients and provides a compact summary of economically meaningful short- and medium-run effects.

\subsection{Effects on Store Survival, Employment, and Earnings}

\begin{figure}[htbp!]
    \centering
    \begin{minipage}{1\linewidth}
    \includegraphics[width=1\linewidth]{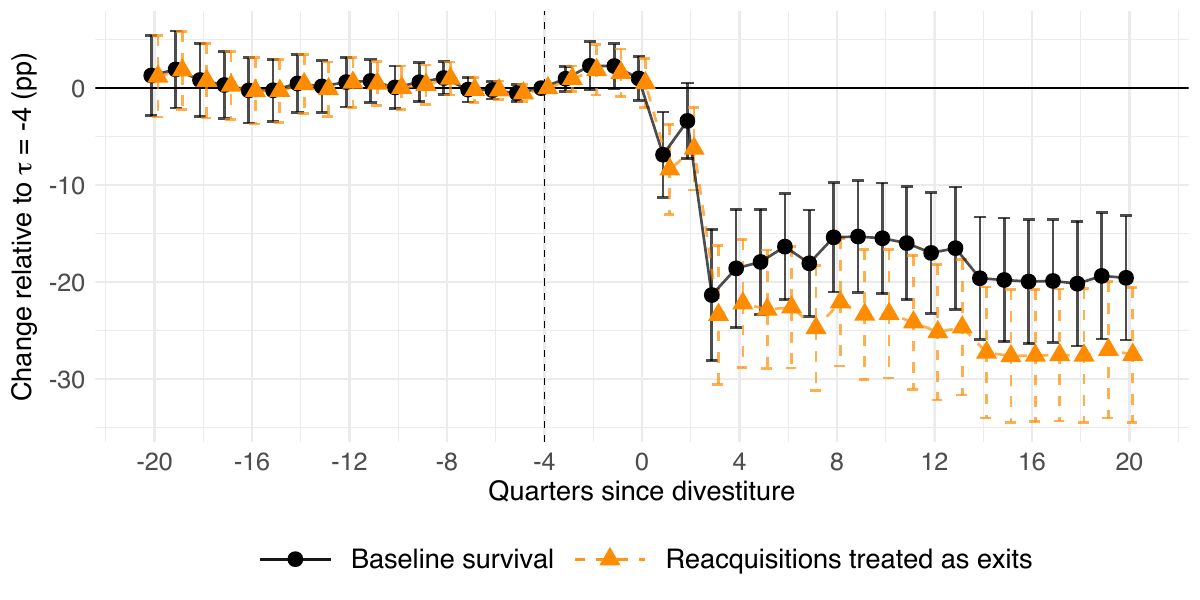}
    \caption{Event-Study Estimates of the Effect of Divestiture on Store Survival}
    \label{fig:survival_event_combined}
    \begin{flushleft}
    \footnotesize
        \textit{Notes:} All estimates are based on data from the BLS QCEW LDB. The figure plots the event-study estimates of the effect of divestiture on store survival. In the ``reacquisitions treated as exit'' case, reacquisitions by the merging parties are counted as exits.
    \end{flushleft}
    \end{minipage}
\end{figure}

\paragraph{Store Survival}
We first examine store survival by estimating specification \eqref{eq:event_study} with an indicator for whether the store is operating as the dependent variable.\footnote{In \figref{fig:survival}, we plot the survival rates of the treated and control groups. Over the six years following the baseline quarter (four quarters prior to divestiture), survival in the control group declines by an average of 2.3 percent per year.} The solid black line in \figref{fig:survival_event_combined} reports the resulting event-study coefficients. We find no evidence of differential pre-trends prior to divestiture. Following divestiture, survival probabilities decline sharply. Within the first three quarters, survival at treated stores falls by up to 21 percentage points relative to controls. This gap then stabilizes: survival remains 19 percentage points lower after one year, 15 percentage points lower after two years, and 20 percentage points lower after five years.

\tabref{table:did_bls} reports the corresponding average treatment effects at 1-, 2-, and 5-year horizons based on equation \eqref{eq:lin_com}, which imply average declines of 13 percentage points after one year, 15 percentage points after two years, and 17 percentage points after five years. Taken together, these estimates indicate that divested establishments are substantially more likely to exit in the post-divestiture period.

\begin{table}[htbp]
\small
\centering
\caption{Horizon-Specific Average Effects of Divestiture}
\label{table:did_bls}
\begin{threeparttable}
\begin{tabular}{lccccc}
\toprule
 & Survival (pp) & Employment (\%) & DHS Growth Rate (pp) & Payroll (\%) & Avg. Wage (\%)  \\
Post Horizon & (1) & (2) & (3) & (4) & (5) \\ \midrule
1 Year & -12.5$^{***}$ & -22.4$^{***}$ & -47.5$^{***}$ & -21.5$^{***}$ & 1.2 \\
 & (2.3) & (2.8) & (4.6) & (2.6) & (2.1) \\
2 Years & -14.7$^{***}$ & -18.9$^{***}$ & -49.8$^{***}$ & -21.9$^{***}$ & -3.7$^{*}$ \\
 & (2.4) & (2.4) & (4.3) & (2.5) & (1.8) \\
5 Years & -16.8$^{***}$ & -15.3$^{***}$ & -52.5$^{***}$ & -21.1$^{***}$ & -6.8$^{***}$ \\
 & (2.4) & (2.8) & (4.5) & (2.8) & (2.0) \\
\bottomrule
\end{tabular}
    \begin{tablenotes}
        \footnotesize
        \item \emph{Notes:} *** $p<0.01$, ** $p<0.05$, * $p < 0.10$. The table reports the horizon-specific average treatment effects calculated via equation \eqref{eq:lin_com}. Control stores are defined as described in \secref{sub:control}. All estimates are based on data from the BLS QCEW LDB. Survival and DHS Growth Rate are reported in percentage-point units. Employment, Payroll, and Average Wage are reported in percentage-change units, obtained by transforming log-change estimates as $100(\exp(\hat\beta)-1)$; standard errors for these outcomes are updated using the delta method.
    \end{tablenotes}
\end{threeparttable}
\end{table}

One concern in using store survival to assess the effectiveness of divestiture remedies is that some divested stores may later be reacquired by the merging parties. In such cases, measured survival would overstate the effectiveness of divestiture as a competitive remedy. To address this concern, we identify reacquired stores as those that transition to an EIN associated with the merging firms and re-estimate the survival specification, treating these stores as exits. The orange line in \figref{fig:survival_event_combined} reports the resulting event-study coefficients. Under this classification, survival declines by 22 percentage points after one year and after two years, and 27 percentage points after five years relative to controls. These results indicate that reacquisitions contribute meaningfully to observed post-divestiture survival.

\paragraph{Store Employment}
We next examine the effects of divestiture on employment by estimating equation \eqref{eq:event_study} using log employment as the dependent variable. Because survival effects are substantial, we condition the event-study analysis on establishments that remain in operation 20 quarters after divestiture. By doing so, we isolate intensive-margin adjustments among continuing stores and avoid mechanically conflating exit with employment reductions.\footnote{If a store temporarily closes and later reopens, we exclude both the treated store and its associated controls during the period in which the store is closed.}

\begin{figure}[htbp!]
    \centering
    \includegraphics[width=1\linewidth]{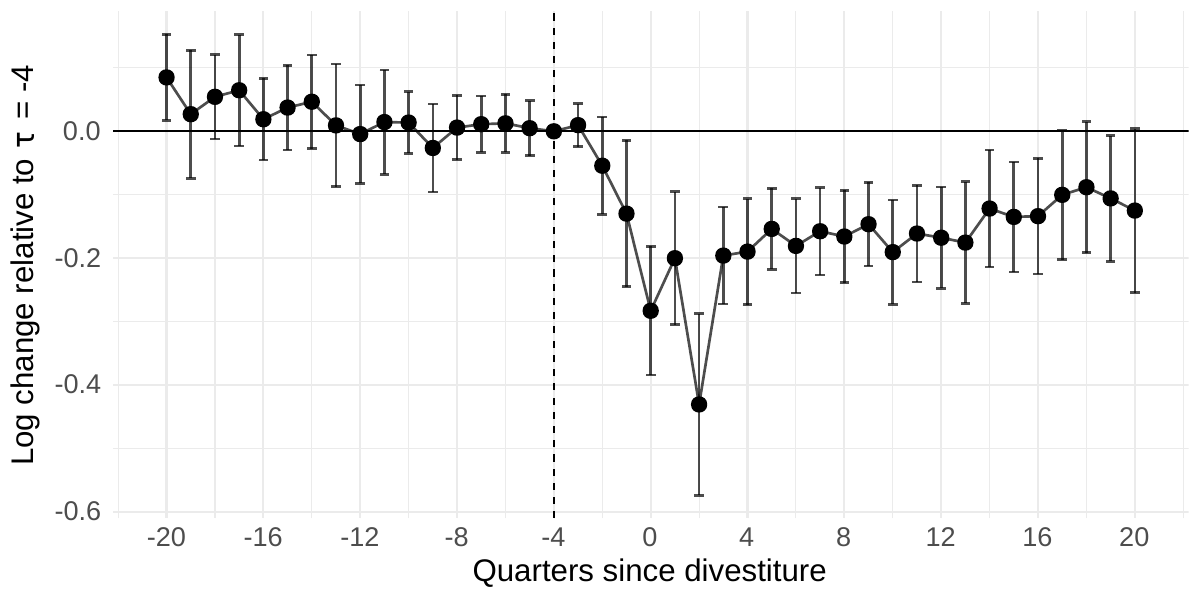}
    \caption{Event-Study Estimates of the Effect of Divestiture on Surviving Store Employment}\label{fig:emp_event}
    \begin{flushleft}
        \footnotesize
        \textit{Notes:} All estimates are based on data from the BLS QCEW LDB. The figure plots the estimated percentage change in store employment for treated stores relative to control stores around the time of the divestiture, restricting the sample to stores that survive for 20 quarters. The reported percentage changes are obtained by transforming log changes using $100 \times (e^{\hat{\beta}}-1)$.
    \end{flushleft}
\end{figure}

\figref{fig:emp_event} reports the estimated event-study coefficients, expressed in percentage changes relative to four quarters prior to divestiture. We find no evidence of differential trends prior to $\tau = -4$. However, employment declines sharply in the quarters immediately preceding and surrounding divestiture: employment at treated stores is 5 percent lower two quarters prior to divestiture, 12 percent lower one quarter prior, and 25 percent lower in the quarter of divestiture relative to controls. These short-run declines likely reflect transition dynamics associated with ownership transfer.

Employment continues to fall in the immediate post-divestiture period, reaching a decline of up to 35 percent two quarters after divestiture, before stabilizing. Over longer horizons, employment remains persistently lower at divested stores: 17 percent lower after one year, 15 percent lower after two years, and 13 percent lower after five years relative to controls. As reported in \tabref{table:did_bls}, the average employment effect equals -22 percent over the first year, -19 percent over two years, and -15 percent over five years.\footnote{The reported percentage changes are obtained by transforming log changes using $100 \times (e^{\hat{\beta}}-1)$.}

\paragraph{Total Employment Decomposition}
Taken together, the preceding results indicate that divestiture reduces both the extensive margin of store survival and the intensive margin of employment among surviving stores. To quantify the total effect on employment, we exploit the identity
\begin{equation}\label{eqn:employment.identity}
E(\mathit{Emp}_{it}) = E(\mathit{Emp}_{it} | \mathit{Survival}_{it} = 1) P(\mathit{Survival}_{it} = 1).
\end{equation}
Equation \eqref{eqn:employment.identity} implies that the log change in expected employment following a divestiture equals the sum of the log change in employment among surviving stores and the log change in survival probabilities:
\begin{equation}\label{eqn:employment.identity.divestiture}
\frac{\partial \log E\left( \mathit{Emp}_{it} \right) }
{\partial D_{it}}
=
\frac{\partial \log E\left( \mathit{Emp}_{it} \mid \mathit{Survival}_{it}=1 \right)}
{\partial D_{it}}
+
\frac{\partial \log  P\left( \mathit{Survival}_{it}=1 \right) }
{\partial D_{it}},
\end{equation}
where $D_{it}$ is the divestiture indicator.

We estimate the two components on the right-hand side of \eqref{eqn:employment.identity.divestiture} using the estimates reported in \tabref{table:did_bls}. The intensive-margin component is obtained from Column 2, which reports estimates from an event-study regression with log employment as the outcome.\footnote{Because the regression estimates 
$E[\log(\mathit{Emp}_{it}) \mid \mathit{Survival}_{it}=1]$ rather than $\log E[\mathit{Emp}_{it} \mid \mathit{Survival}_{it}=1]$, 
the decomposition is exact under a log-linear model with additive errors independent of treatment. More generally, it should be interpreted as a first-order approximation to the effect on log mean employment.} The extensive-margin component is constructed from the survival estimates in Column 1 by converting the estimated percentage-point effect on survival into a log change using the
control-group survival rate.\footnote{Formally, the implied log effect in
\eqref{eqn:employment.identity.divestiture} equals
$\beta_e + \log\!\big((P_c+\beta_s)/P_c\big)$, where $\beta_e$ denotes the estimated effect of treatment on log employment conditional on survival, $\beta_s$ denotes the estimated
percentage-point effect on the probability of survival, and $P_c$ is the average survival rate in the control group.} Using this decomposition, we estimate that divestiture reduces total employment by approximately 33 percent after one year and by about 31 percent after both two and five years. These magnitudes are consistent with the patterns in \figref{fig:employment} and indicate that roughly one-half of the overall employment decline is attributable to intensive-margin reductions among surviving stores.

As an alternative approach that jointly captures intensive- and extensive-margin employment adjustments, we use the symmetric growth rate measure of \citet{davis1998job}. This measure allows employment changes and store exit to be analyzed within a single outcome variable, avoiding the need to condition on survival. In equation \eqref{eq:event_study}, we replace $Y_{it}$ with
\begin{equation} \label{eqn:dhs}
g_{it} = \frac{\mathit{Emp}_{it} - \mathit{Emp}_{i,\tau=-4}}{\frac{1}{2}(\mathit{Emp}_{it} + \mathit{Emp}_{i,\tau=-4})}.
\end{equation}

This measure is approximately equal to log employment growth for small changes and remains well-defined in the presence of entry and exit; in particular, $g_{it} = -2$ for firms that exit (i.e., when $\mathit{Emp}_{it}=0$). Because it uses the average of employment in both periods in the denominator, the symmetric growth rate treats expansions and contractions symmetrically and permits aggregation consistent with the Davis–Haltiwanger–Schuh decomposition of job creation and destruction. Column 3 of \tabref{table:did_bls} reports estimates using this outcome. We find declines in the symmetric growth rate of 48 percentage points after one year, 50 percentage points after two years, and 53 percentage points after five years.\footnote{The symmetric growth rate estimates and the log-decomposition results are different because the symmetric growth rate applies a nonlinear transformation to employment at the store level before aggregation, whereas the decomposition combines separately estimated margin effects under a log-linear approximation.}

\paragraph{Store Payroll and Average Earnings}

The QCEW LDB data also provides quarterly payroll for each establishment. We estimate equation \eqref{eq:event_study} using log payroll as the dependent variable, conditioning on establishments that remain in operation 20 quarters after divestiture as in the employment analysis earlier.

\begin{figure}[htbp!]
    \centering
    \begin{minipage}{1\linewidth}
    \begin{subfigure}[b]{1\linewidth}
        \centering
        \includegraphics[width=\linewidth]{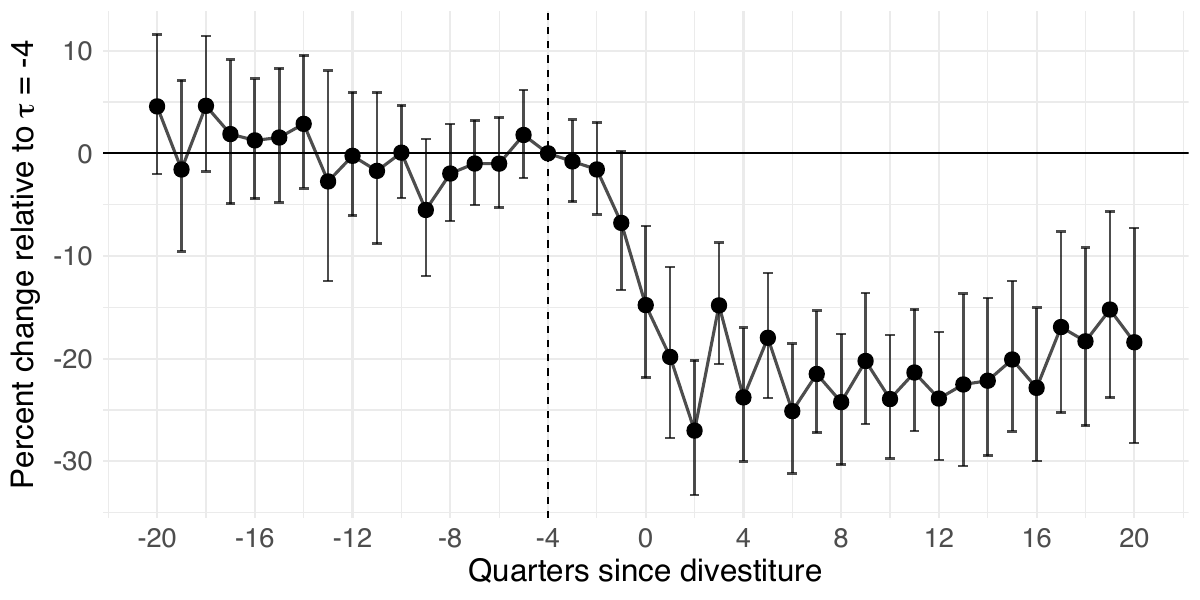}
        \caption{Payroll}
        \label{fig:payroll_event}
    \end{subfigure}\hfill
    \begin{subfigure}[b]{1\linewidth}
        \centering
        \includegraphics[width=\linewidth]{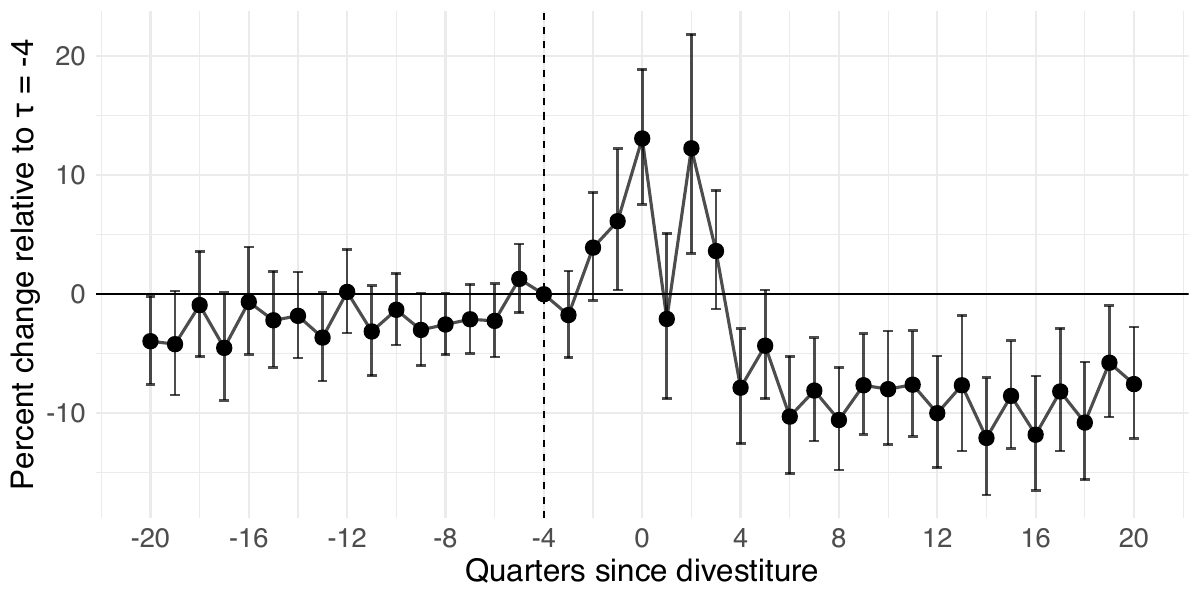}
        \caption{Average Earnings}
        \label{fig:earnings_event}
    \end{subfigure}
    \caption{Event-Study Estimates of the Effect of Divestiture on Store Payroll and Average Earnings}\label{fig:events}
    \begin{flushleft}
        \footnotesize
        \textit{Notes:} All estimates are based on data from the BLS QCEW LDB. The figure plots the log change in payroll (panel (a)) and average earnings (panel (b)) for the treatment and control groups around the time of the divestiture, restricting the sample to only stores that survive 20 quarters. Average earnings is measured as payroll divided by employment. The reported percentage changes are obtained by transforming log changes using $100 \times (e^{\hat{\beta}}-1)$.
    \end{flushleft}
    \end{minipage}
\end{figure}

\figref{fig:payroll_event} presents the event-study estimates. Relative to the four quarters prior to divestiture, we find no evidence of differential payroll trends prior to treatment. Payroll declines sharply around the time of divestiture, falling by approximately 7 percent in the quarter prior and by 16 percent in the quarter of divestiture relative to control stores. The decline deepens in the immediate post-divestiture period. Over longer horizons, payroll remains persistently lower at divested stores, declining by 27 percent after one year and 28 percent after two years, and by 20 percent after five years relative to controls. The corresponding horizon-averaged effects reported in \tabref{table:did_bls} imply an average reduction between 21 and 22 percent over a one to five-year horizon.

Because payroll contracts more than employment, divestiture also affects average earnings. \figref{fig:earnings_event} reports event-study estimates for log average earnings, again conditioning on survival to 20 quarters. In earlier pre-divestiture quarters, average earnings at treated stores are modestly below the baseline quarter of $\tau = -4$, but we do not observe systematic divergence in the quarters immediately preceding divestiture. Around the time of divestiture, average earnings temporarily increase, reaching as much as 13 percent above baseline in some quarters. This short-run rise coincides with sharp declines in employment and is consistent with compositional adjustments in the workforce following the ownership transfer.

Beginning approximately four quarters after divestiture, however, average earnings decline persistently relative to control stores. Average earnings are 8 percent lower one year post-divestiture, 11 percent lower after two years, and 8 percent lower after five years. The horizon-averaged estimates indicate no statistically significant change over the first year but declines of approximately 4 percent over two years and 7 percent over five years. Taken together, these results indicate that divested stores both reduce employment and modestly reduce compensation among remaining workers in the short and medium run.

\paragraph{Divestiture Effects Across Time}

Finally, we examine whether the effects of divestiture vary across enforcement regimes. The FTC’s first divestiture retrospective \citep{ftc1999divestiture} prompted a series of reforms intended to improve buyer quality and reduce the risk that divested assets would fail. In particular, the FTC increasingly required an upfront buyer when the divestiture did not involve the sale of a complete ongoing business, shortened the period allowed for completing non-upfront divestitures, and implemented post-divestiture follow-up interviews with buyers to assess implementation.

\begin{figure}[htbp!]
    \centering
    \begin{minipage}{1\textwidth}
    \begin{subfigure}[b]{1\textwidth}
        \centering
        \includegraphics[width=\textwidth]{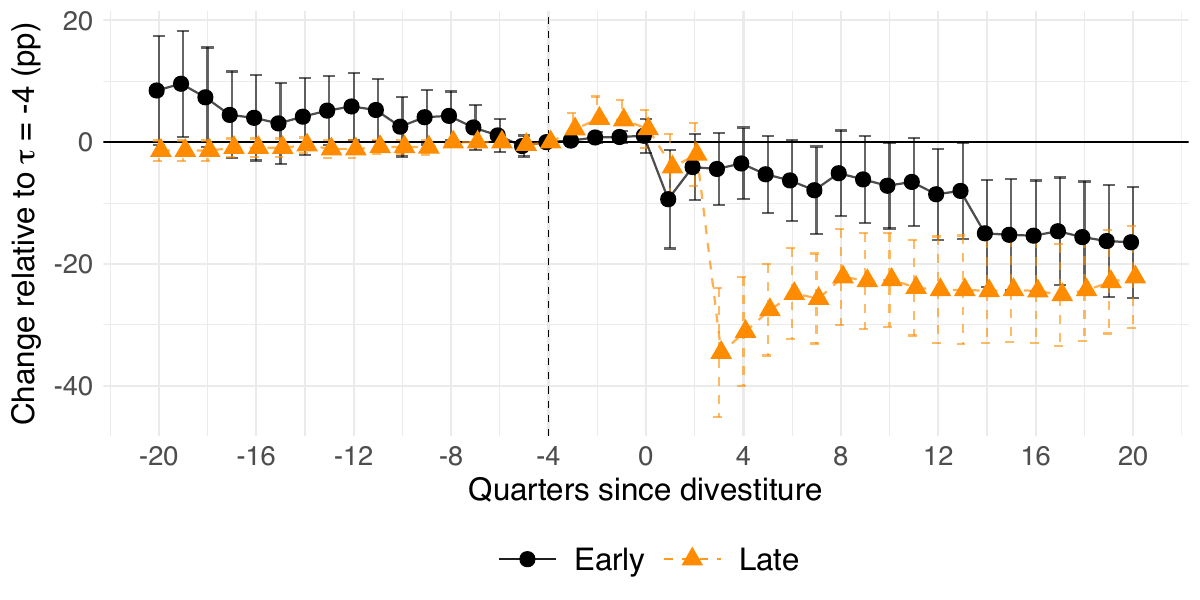}
        \caption{Survival Rate}
        \label{fig:sub2}
    \end{subfigure}
        \hfill
    \begin{subfigure}[b]{1\textwidth}
        \centering
        \includegraphics[width=\textwidth]{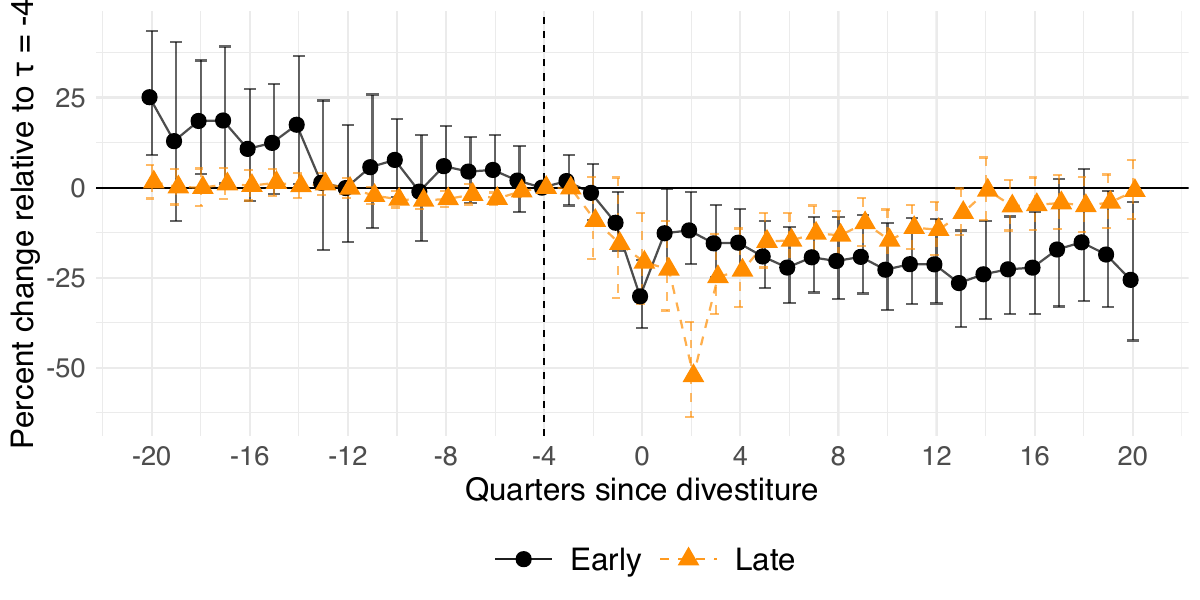}
        \caption{Employment}
    \end{subfigure}
    \caption{Survival and Employment Pre- and Post-2008}\label{fig:early_late}
    \begin{flushleft}
        \footnotesize
        \textit{Notes:} All estimates are based on data from the BLS QCEW LDB. The figures plots the event-study estimates of the effect of divestiture on store survival rates (panel (a)) and employment (panel (b)). ``Early'' and ``Late'' correspond to the pre-2008 and post-2008 periods, respectively. The reported percentage changes in panel (b) are obtained by transforming log changes using $100 \times (e^{\hat{\beta}}-1)$.
    \end{flushleft}
    \end{minipage}
\end{figure}

To assess whether outcomes improved following these changes, we partition transactions into those occurring before January 2008 and those thereafter. This cutoff roughly splits the sample in half: the early period primarily consists of mergers from the 1990s, while the later period consists mostly of mergers from the 2010s. By the later period, upfront-buyer provisions had become standard in supermarket merger remedies. \figref{fig:early_late} reports event-study estimates for survival (top panel) and employment among surviving stores (bottom panel), with the early period shown as a solid black line and the later period as a dashed orange line.

For survival, the post-divestiture dynamics differ across periods. In the first year after divestiture, survival declines by 3 percentage points in the early period, compared to 31 percentage points in the later period. However, the early-period effect worsens over time, while the later-period effect attenuates. As a result, the survival effects partially converge, reaching approximately 16 percentage points after five years for the early period and 22 percentage points for the late period.

Employment dynamics also differ across periods. In the early period, employment among surviving stores declines progressively over time, from a 15 percent decline after one year to a 26 percent decline after five years. In addition, the early subsample exhibits a downward pre-divestiture trend relative to the baseline quarter ($\tau = -4$); employment at treated stores is approximately 25 percent higher five years prior to divestiture than at $\tau = -4$, indicating that employment was already declining leading up to divestiture. This pattern suggests that some early divestitures may have occurred in stores already experiencing declining performance.

In contrast, the later period shows no evidence of differential pre-trends between treated and control stores. Employment declines sharply immediately following divestiture, falling by 23 percent after one year, but partially recovers thereafter, with only a 13 percent decline after two years and a negligible difference after five years.

To compare total employment effects, we apply the same decomposition framework as in equation \eqref{eqn:employment.identity.divestiture}. In the first year, total employment declines by 19 percent in the early period, substantially smaller than the 45 percent decline in the later period. Over two years, the difference persists, with a 22 percent decline in the early period compared to a 41 percent decline in the later period. On average over five years, however, the magnitude of the total employment effect increases to a 28 percent decline for the early period, compared to a 35 percent decline for the later period. These results suggest that the reforms did not materially change the long-run employment decline for divested stores in the supermarket industry.

\subsection{Effects on Sales and Profitability}

We next examine store-level revenues and profit margins to assess whether divested stores operate at reduced scale or profitability following divestiture. Because the BLS data do not contain revenue or margin measures, we analyze proprietary data obtained directly from supermarket chains by the FTC, which include store-level sales and operating income. As these data are available for specific transactions rather than the full universe of divestitures, we analyze each event separately using a difference-in-differences framework analogous to the earlier event-study specification.\footnote{Since each divestiture affects all treated stores at a common time (i.e., there is no staggered adoption), we use the standard event-study specification $Y_{it} = \alpha_i + \gamma_t + \sum_{k \neq k_0} \beta_k \mathbf{1}\{t-t_0 = k\} \cdot D_i + \varepsilon_{it}$, where event-time effects are measured relative to the reference period $t-t_0 = k_0$ (e.g., $k_0 = -1$). }

\begin{figure}[htbp]
    \centering
    \begin{minipage}{1\textwidth}
    \begin{subfigure}[t]{0.48\textwidth}
        \centering
        \includegraphics[width=\textwidth]{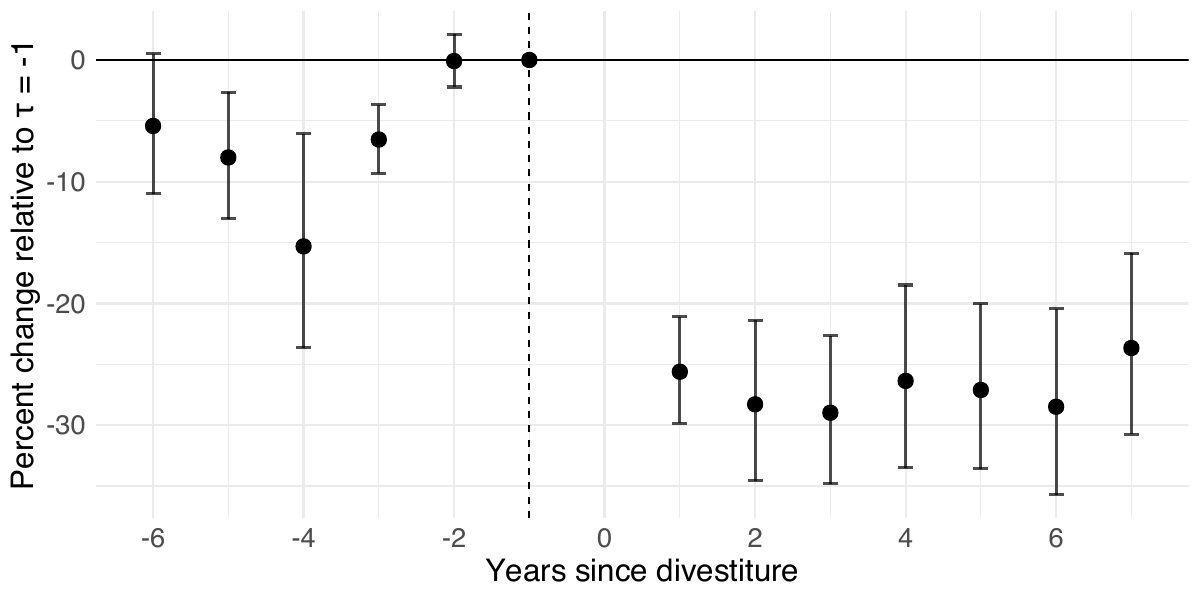}
        \caption{Divestiture Event 1}
        \label{figure:Div1LogSalesEventStudy}
    \end{subfigure}%
    \hfill
    \begin{subfigure}[t]{0.48\textwidth}
        \centering
        \includegraphics[width=\textwidth]{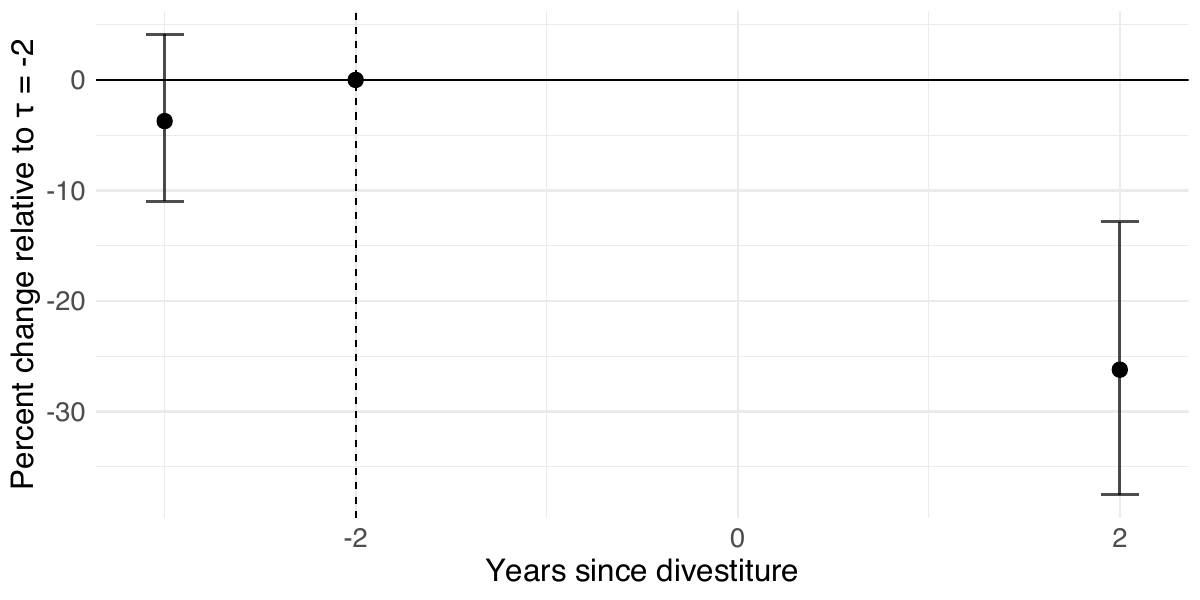}
        \caption{Divestiture Event 2}
        \label{figure:Div2LogSalesEventStudy}
    \end{subfigure}
    \caption{Event-Study Estimates of Divestiture Effects on Store Sales}
    \label{figure:the.impact.of.divestiture.on.store.store.sales}    
    \begin{flushleft}
        \footnotesize
        \textit{Notes:} Calculations are based on analysis of confidential investigation data from the FTC. The figures plot the event-study estimates of the effects of divestiture on log store sales for two anonymous divestiture events. To protect confidentiality, the identities of the divestitures are not disclosed. The reported percentage changes are obtained by transforming log changes using $100 \times (e^{\hat{\beta}}-1)$.
    \end{flushleft}
    \end{minipage}
\end{figure}

For the first divestiture event, we consider three sets of control groups: (i) stores belonging to unrelated chains, (ii) incumbent stores of the divestiture buyer, and (iii) non-divested stores of the divestiture seller. \figref{figure:Div1LogSalesEventStudy} presents event-study estimates pooling these controls. Sales decline sharply at the time of divestiture and show little evidence of recovery over the subsequent years. The event-study coefficients imply a reduction of approximately 27 percent relative to the year preceding divestiture. \tabref{table:did_sales} reports the average of these estimates across the post-treatment window, calculated via \eqref{eq:lin_com}. Across alternative control groups (columns (1) to (4)), the estimated decline in store-level sales ranges from 18 percent when using incumbent buyer stores as controls to 31 percent when using non-divested seller stores.\footnote{The composition and time coverage of the control groups vary due to data availability. As a result, the pooled control group in column (4) forms an unbalanced panel, whereas the panels used in columns (1)–(3) are relatively balanced.}

\begin{table}[htbp]
\small
\centering
\caption{Effect of Divestiture on Sales (\%)}
\label{table:did_sales}
\begin{threeparttable}
\begin{tabular}{lcccccc}
\toprule
 & \multicolumn{4}{c}{Divestiture 1} & \multicolumn{2}{c}{Divestiture 2} \\ \cmidrule(lr){2-5} \cmidrule(lr){6-7}
 & Unrelated Chains & Buyer & Seller  & All & Full Years & All  \\   
 & (1) & (2) & (3) & (4) & (5) & (6) \\ \midrule
 Divested & -19.7$^{***}$ & -17.6$^{***}$ & -30.9$^{***}$ & -27.0$^{***}$ & -26.2$^{**}$ & -30.4$^{***}$ \\
 Indicator & (4.3) & (4.2) & (2.8)  & (2.9) & (6.3) & (5.3) \\   
 \midrule 
 $R^2$  & 0.857 & 0.889 & 0.942 & 0.911 & 0.932 & 0.834 \\   
\bottomrule
\end{tabular}
    \begin{tablenotes}
        \footnotesize
        \item \emph{Notes:} *** $p<0.01$, ** $p<0.05$, * $p < 0.10$. The dependent variable is the log of sales excluding fuel and pharmacy. Control stores are located in the same state as treated stores but in different three-digit ZIP codes; for Divestiture 2, we also include stores of a neighboring state in the control group. For both Divestitures 1 and 2, we also exclude the year of the divestiture ($t = 0$) due to partial-year data reporting. For Divestiture 2, we use $t = -2$ as the reference period. All reported coefficients are converted to percentage changes using $100(\exp(\hat\beta)-1)$. Standard errors are transformed using the delta method, $100\exp(\hat\beta)\cdot se(\hat\beta)$.
    \end{tablenotes}
    \end{threeparttable}
\end{table}

We next turn to Divestiture 2. Owing to data limitations, we restrict the control group to stores from an alternative chain and include stores located in a neighboring state. Because sales data for some years around the divestiture reflect only a subset of months, we estimate two specifications. The first restricts the sample to years with complete annual data. The second includes all available years, including those with partial reporting, and scales partial-year sales to an annual frequency.\footnote{For Divestiture 2, treatment-group sales are only partially observed in the years immediately before and after the divestiture ($\tau \in {-1,1}$), and no observations are available in the divestiture year ($\tau = 0$). When partial-year data are used, we scale reported sales to an annual equivalent. For example, if only 28 weeks of sales are observed in $\tau = -1$, we multiply sales by $52/28$.} In both specifications, we use $\tau = -2$ as the reference period because sales in $\tau = -1$ are only partially observed.

\figref{figure:Div2LogSalesEventStudy} shows that sales decline by approximately 26 percent in the second year following divestiture relative to the pre-divestiture baseline. Columns (5) and (6) of \tabref{table:did_sales} report the average post-divestiture event-study estimates.\footnote{Column (5) (``Full Years'') excludes years with partial reporting. Column (6) (``All'') retains these years and scales sales to annual equivalents using a constant adjustment factor. } The estimates confirm a similar reduction of roughly 26--30 percent across alternative sample restrictions.

\begin{figure}[htbpt]
    \centering
    \begin{minipage}{1\textwidth}
    \begin{subfigure}[t]{0.48\textwidth}
        \centering
        \includegraphics[width=\textwidth]{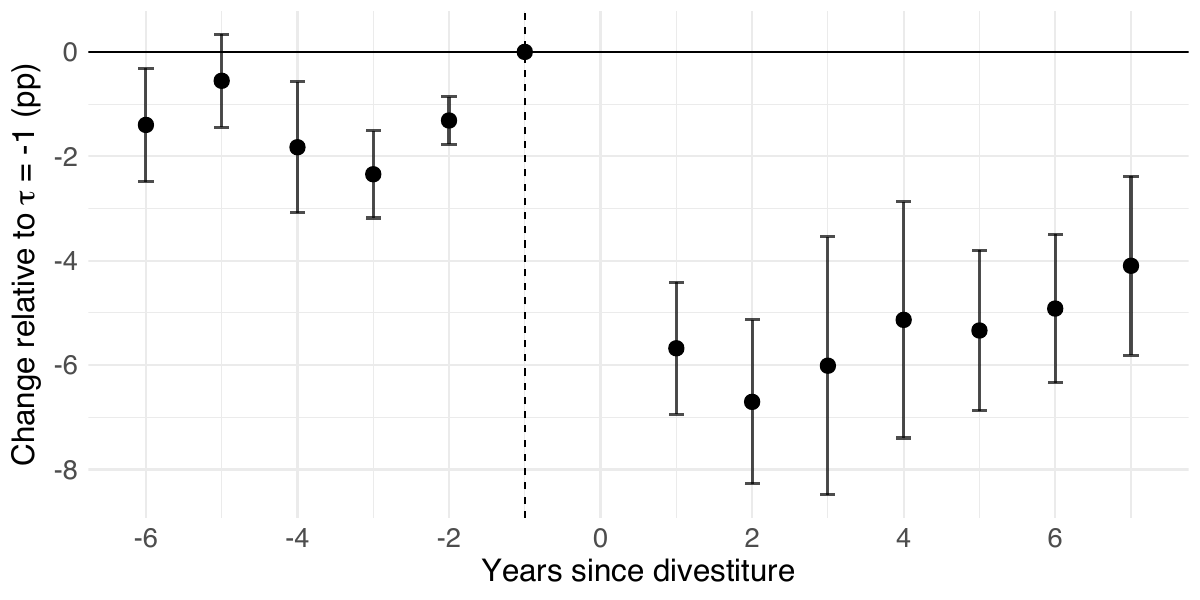}
        \caption{Divestiture Event 1}
        \label{figure:Div1NetMarginsEventStudy}
    \end{subfigure}%
    \hfill
    \begin{subfigure}[t]{0.48\textwidth}
        \centering
        \includegraphics[width=\textwidth]{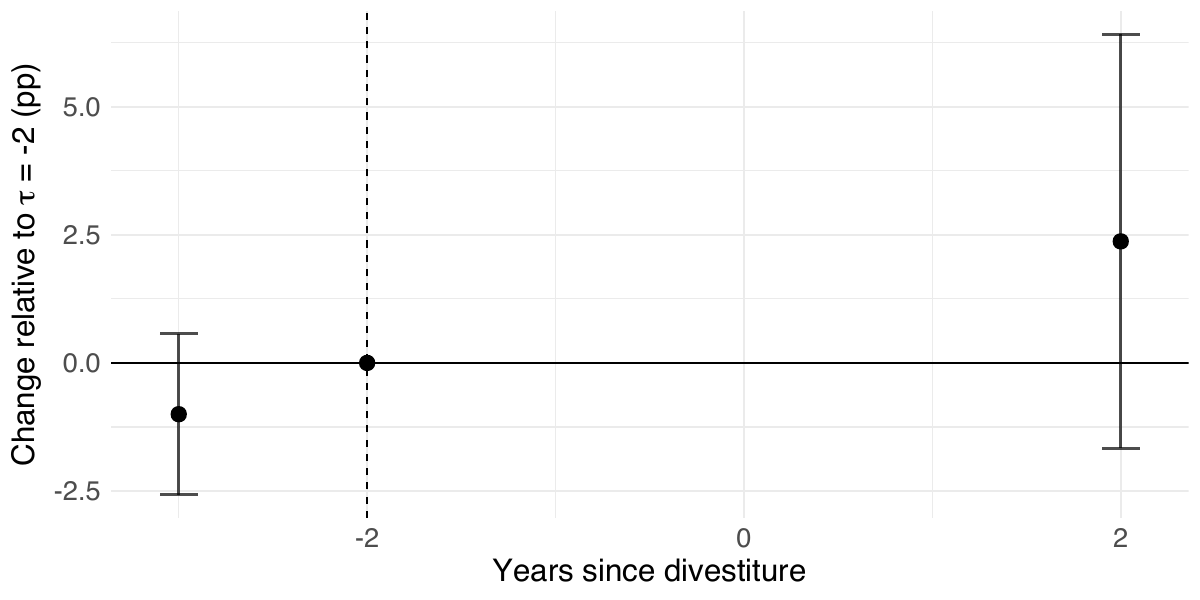}
        \caption{Divestiture Event 2}
        \label{figure:Div2NetMarginsEventStudy}
    \end{subfigure}
    \caption{Event-Study Estimates of Divestiture Effects on Store Net Margins}
    \label{figure:the.impact.of.divestiture.on.net.margins}
    \begin{flushleft}
        \footnotesize
        \textit{Notes:} Calculations are based on analysis of confidential investigation data from the FTC. The figures plot the event-study estimates of the effects of divestiture on store net margins for two anonymous divestiture events. To protect confidentiality, the identities of the divestitures are not disclosed.
    \end{flushleft}
    \end{minipage}
    
\end{figure}

We next examine store-level profitability using net profit margins, defined as operating income divided by sales. \figref{figure:the.impact.of.divestiture.on.net.margins} and \tabref{table:did_net_margin} report the event-study estimates and their post-divestiture averages analogous to those presented for sales.

\begin{table}[htbp]
\small
    \centering
    \caption{Effect of Divestiture on Store Net Margins (p.p.)}
    \label{table:did_net_margin}

\centering
\begin{threeparttable}
\begin{tabular}{lccccc}
\toprule
 & \multicolumn{3}{c}{Divestiture 1}  & \multicolumn{2}{c}{Divestiture 2} \\ \cmidrule(lr){2-4} \cmidrule(lr){5-6}
 & Unrelated Chains & Divestiture Buyer & All & Full Years & All \\   
 & (1) & (2) & (3) & (4) & (5) \\ \midrule
 Divested  & -6.3$^{***}$ & -5.4$^{***}$ & -5.4$^{***}$ & 2.4 & -0.5\\  
 Indicator & (0.9) & (1.3)  & (0.7) & (2.1) & (1.8) \\    \midrule
   $R^2$ & 0.769 & 0.808 & 0.769 & 0.796 & 0.717 \\   
\bottomrule
\end{tabular}
    \begin{tablenotes}
        \footnotesize
        \item \emph{Notes:} *** $p<0.01$, ** $p<0.05$, * $p < 0.10$. The dependent variable is the net profit margin. Control stores are located in the same state as treated stores but in different three-digit ZIP codes; for Divestiture 2, we also include stores of a neighboring state in the control group. For both Divestitures 1 and 2, we also exclude the year of the divestiture ($t = 0$) due to partial-year data reporting. For Divestiture 2, we use $t = -2$ as the reference period. All coefficients are reported in percentage-point (PP) units.
    \end{tablenotes}
    \end{threeparttable}
\end{table}

For Divestiture Event 1, we document a persistent decline in store-level net margins following divestiture. The estimated effects imply a reduction of approximately 5--6 percentage points, and the magnitude is stable across alternative control groups.\footnote{Net margin information is unavailable for the control group consisting of non-divested stores owned by the divestiture seller. Accordingly, the analysis relies on control stores from other chains and incumbent buyer stores.} Given an average pre-divestiture margin of 6.0 percent, this decline corresponds to a reduction in profitability of more than 80 percent. In an industry characterized by thin margins, such a contraction represents a substantial deterioration in store performance.

\tabref{table:did_net_margin} columns (4)--(5) report the corresponding estimates for Divestiture Event 2. In contrast to Divestiture Event 1, the effects are smaller and imprecisely estimated. Using the full sample (column (5)) yields a statistically insignificant decline of less than one percentage point. Restricting the sample to years with complete data (column (4)) produces a positive point estimate of 2.4 percentage points, which is likewise statistically insignificant. We therefore find no robust evidence of either margin improvement or deterioration following this divestiture. Taken together, the sales and margin results indicate that divested stores operate at materially reduced scale and, in at least one case, experience a pronounced and economically meaningful decline in profitability.

%% file: Mechanism.tex
\section{Determinants of Post-Divestiture Performance \label{section:mechanism}}

Merging firms typically negotiate with antitrust authorities over the scope of assets to be divested, the identity of the buyer, and, in some cases, additional terms governing the remedy. Because the merged firm continues to compete against the divested assets, its incentives may not align with the regulator’s objective of preserving competition. Firms may therefore propose divestiture packages that involve relatively weaker assets, less capable buyers, or narrowly scoped remedies, risks that prior FTC reports and statements have emphasized \citep{ftc1999divestiture, FTC2012negotiating, ftc2017merger, parker2000evolving}. In this section, we examine three economic mechanisms that may account for the observed post-divestiture performance: asset quality, buyer quality, and transition dynamics. We perform each analysis using data available from the FTC which are only available for a subset of mergers.

\subsection{Asset Quality}

The effectiveness of a divestiture may depend on the quality of the assets transferred. In retail mergers, where store locations are the primary assets, stores differ substantially in profitability, customer base, and competitive positioning. Because the merged firm continues to compete against the divested assets, its incentives may not align with the regulator’s objective of preserving competitive strength; instead, it may want to retain relatively stronger stores and divest weaker ones. We therefore examine whether divested stores differ systematically from retained stores along pre-merger performance measures.

To assess whether merging firms divest systematically weaker assets, we analyzed proprietary store-level data from two prior grocery divestitures (Divestitures 1 and 2).\footnote{The divestitures labeled ``Divestiture 1'' and ``Divestiture 2'' in this subsection do not necessarily correspond to those labeled with the same names in the previous section. The same caveat applies to subsequent sections.} We compare the pre-merger performance of divested stores to that of nearby retained stores within a 10-mile radius, focusing on the calendar year immediately preceding the merger. \tabref{table:retained_divested} reports the percentage difference in average annual sales, sales per square foot, and profit margins measured as the EBITDA to sales ratio.

\begin{table}[htbp!]
\small
    \centering
    \caption{Pre-Merger Performance of Divested Stores Relative to Retained Stores}
    \label{table:retained_divested}
    \begin{threeparttable}
    \begin{tabular}{lccc} \toprule
    & \multicolumn{3}{c}{Measures of Financial Performance} \\ \cmidrule(lr){2-4}
    & Annual Sales & Annual Sales/Sqft & Profit Margin \\ \midrule
    Divestiture 1 & -17.3\% & -17.2\% & -44.4\%\\ 
    Divestiture 2 & -61.5\% & -42.1\% & -90.1\% \\  \bottomrule
    \end{tabular}
    \begin{tablenotes}
        \footnotesize
        \item \emph{Notes:} The table reports percentage differences in average pre-merger financial performance between divested and retained stores for two grocery divestitures. Outcomes are measured in the year immediately preceding the merger. Sales per square foot are computed using selling space (in square feet). Profit margin is defined as EBITDA divided by sales. The relative difference is calculated as $(\bar{Y}_{\mathit{Divested}} - \bar{Y}_{\mathit{Retained}})/\bar{Y}_{\mathit{Retained}}$.
    \end{tablenotes}
    \end{threeparttable}
\end{table}

In both divestitures, divested stores were substantially weaker than nearby retained stores prior to the merger. In Divestiture 1, divested stores generated 17 percent lower sales and sales per square foot and 44 percent lower profit margins than retained stores. In Divestiture 2, the gaps are considerably larger: divested stores exhibit 62 percent lower sales, 42 percent lower sales per square foot, and 90 percent lower profit margins than retained stores. In the latter case, divested stores appear close to break-even profitability before divestiture.

These patterns suggest that divestitures disproportionately involve lower-performing stores within the seller’s portfolio. Because asset quality influences competitive viability, transferring weaker stores may limit the buyer’s ability to replicate the competitive constraint of the pre-merger firm. This selection is consistent with the incentive concerns emphasized in prior FTC analyses, which note that merging firms may structure divestiture packages in ways that attenuate the competitive strength of the remedy.%

\subsection{Buyer Quality}

Buyer quality is likely an important determinant of post-divestiture performance. Buyers with greater scale, established supply chains, and brand recognition may operate divested stores more efficiently, whereas smaller or less-experienced buyers may have higher costs or face integration challenges. Poor post-divestiture outcomes may therefore reflect limitations of the acquiring firm rather than weaknesses of the divested assets themselves.

\paragraph{Relative Buyer Performance}

To assess buyer quality, we compare the pre-divestiture performance of the buyer’s existing stores to that of the merged firm’s stores using data from two divestiture events. As in the asset-quality analysis, we examine percentage differences in average sales, sales per square foot, and profit margins in the year preceding the transaction. \tabref{table:seller_buyer} reports the results.

\begin{table}[htbp!]
\small
\centering
\begin{threeparttable}

\caption{Pre-Merger Performance of Buyer Stores Relative to Seller Stores}
\label{table:seller_buyer}
\begin{tabular}{lccc} \toprule
& \multicolumn{3}{c}{Measures of Financial Performance} \\ \cmidrule(lr){2-4}
& Annual Sales & Annual Sales/Sqft & Profit Margin \\ \midrule
Divestiture 1 & -15.2\% & -17.1\% & 103.4\% \\
Divestiture 2 & -57.8\% &  -23.6\% & -  \\ \bottomrule
\end{tabular}
       \begin{tablenotes}
        \footnotesize
        \item \emph{Notes:} The table reports percentage differences in average pre-merger financial performance between seller and buyer stores for two grocery transactions. Outcomes are measured in the year immediately preceding the merger. Sales per square foot are computed using selling space (in square feet); square footage is calculated as the firm's regional average (i.e., urban, suburban, rural) selling space when unavailable for specific buyer stores in Divestiture 2. Profit margin is defined as EBITDA divided by sales. Profit margin is not reported for buyer stores in Divestiture 2. Seller performance includes divested stores. The relative difference is calculated as $(\bar{Y}_{\mathit{Buyer}} - \bar{Y}_{\mathit{Seller}})/\bar{Y}_{\mathit{Seller}}$.
        
    \end{tablenotes}
\end{threeparttable}
\end{table}

In both divestitures, the merged firm’s stores exhibit substantially higher sales levels than the buyer’s existing stores. In Divestiture 1, buyer stores generate 15 percent lower annual sales and 17 percent lower sales per square foot than the seller’s stores. In Divestiture 2, the gap is considerably larger: buyer stores exhibit 58 percent lower annual sales and 24\% lower sales per square foot than seller stores.

Profitability comparisons are more heterogeneous across transactions. In Divestiture 1, buyer stores earn almost double the profit margins of the seller's stores, indicating substantially higher margins at the buyer's locations. For Divestiture 2, profit-margin data for the buyer’s stores are not available. These comparisons suggest substantial variation in buyer quality across transactions, which may contribute to heterogeneity in post-divestiture outcomes.

\paragraph{Post-Divestiture Performance}

We next examine whether post-divestiture outcomes vary with buyer quality. As a proxy for buyer quality, we use the size of the buyer’s store fleet at the time of acquisition. Larger chains may benefit from economies of scale in distribution, advertising, and procurement, and fleet size may also proxy for operational capacity, productivity, and brand recognition.

We test whether divested stores acquired by larger chains exhibit stronger long-run sales performance than those acquired by smaller chains. Focusing on a divestiture involving multiple buyers, we estimate a difference-in-differences specification that compares long-run revenue changes across stores purchased by large- and small-fleet buyers to a control group of stores retained by the merged firm. We classify buyers operating at least 500 stores nationwide as large-fleet chains. Because store-level sales are observed only immediately prior to divestiture and again more than five years later, the analysis captures long-run effects.

\tabref{tab:LR Divest buyer sales} reports estimates from this difference-in-differences specification. Column (1) uses all retained stores of the merged firm as controls. Column (2) excludes retained stores located in the same three-digit ZIP codes as divested stores to mitigate potential contamination from local competitive effects.

\begin{table}[htbp!]
\small
\centering
\caption{Long-run Divestiture Effect on Store Sales by Buyer Fleet Size}
\label{tab:LR Divest buyer sales}
\begin{threeparttable}
\begin{tabular}{lcc} \toprule
 & \multicolumn{2}{c}{Sales (\%)}\\ \cmidrule(lr){2-3}  
 & (1) & (2) \\ \midrule
 Pooled Divestitures & -13.7$^{***}$ & -10.2$^{**}$\\   
 & (4.7) & (4.9) \\ \midrule
 Divest to Small Fleet  & -35.6$^{***}$ & -33.1$^{***}$ \\ 
 & (3.3) & (3.4) \\   
 Divest to Large Fleet & 10.5$^{**}$  & 14.9$^{***}$ \\   
 & (4.3) & (4.9) \\  
   \midrule
 Exclude ZIP3 Overlaps & No & Yes \\ 
 \bottomrule
\end{tabular}
\begin{tablenotes}
\footnotesize
    \item \emph{Notes}: *** $p <$ 0.01, ** $p <$ 0.05, * $p <$  0.1. Standard errors in parentheses are clustered at the ZIP3 level. All models include year and store fixed effects. The first row reports estimates from a specification that pools divestitures across all buyers. The second and third rows report estimates from a specification that interacts the divestiture indicator with indicators for small-fleet and large-fleet buyers. Coefficients are reported as percent changes, computed as $100(\exp(\hat\beta)-1)$; standard errors are obtained via the delta method.
\end{tablenotes}
\end{threeparttable}
\end{table}

The first row reports the average long-run change in sales for all divested stores, pooling those acquired by small- and large-fleet buyers. We estimate sales declines of 10.2 to 13.7 percent relative to control stores. These effects are smaller than the post-divestiture sales declines observed in the broader sample reported in \tabref{table:did_sales}.

This pooled regression conceals a large difference between small-fleet and large-fleet divestiture buyers. Stores operated by small-fleet divestiture buyers perform substantially worse than non-divested stores, with declines in sales between 33.1 and 35.6\% post-divestiture. In contrast, large-fleet divestiture buyers see their sales rise by 10.5 to 14.9\% post-divestiture. These results indicate that chain size, as measured by the store fleet at the time of purchase, is an important correlate of long-run store revenues among divested stores. We interpret these results to be consistent with the conclusion that buyer quality is an important determinant of divestiture success.

\subsection{Transition Dynamics}

Beyond asset and buyer quality, divestitures may also affect store performance through transition dynamics following the transfer of ownership. We distinguish between direct transition costs associated with ownership transfer—such as remodeling, re-bannering, and integration—and longer operational adjustment paths, which may reflect asset quality or buyer capability rather than the mechanics of the transfer itself. Since one major change with any divestiture is a change in the ownership of stores, we examine how sales and profits respond to ownership changes unrelated to divestitures to isolate this channel. In addition, divested stores typically need to be rebannered and renovated, so we examine the magnitude of these costs. Finally, we examine buyers’ expectations about post-transfer performance and how accurately those expectations predict realized outcomes.

\paragraph{Ownership Change}

We first examine ownership changes unrelated to merger divestitures to isolate the effect of ownership turnover itself. In these transactions, both parties voluntarily enter the acquisition, and the acquiring firm retains the store’s existing banner. This setting allows us to separate general transition effects associated with ownership turnover from factors specific to the divestiture process.

Because participation is voluntary, these transactions are less likely to reflect buyer-selection distortions. The acquiring firm pursues a transaction it expects to be profitable, and the seller agrees based on the value of the offer. Although ownership turnover may generate temporary adjustment costs, it may also produce scale or scope efficiencies.

We analyze three such ownership-change events. For each event, stores that change ownership serve as the treatment group, and the buyer’s pre-existing stores serve as controls. Control stores are located in the same state but outside the three-digit ZIP codes of the treated stores. 

\begin{figure}[htbp]
    \centering
    \begin{minipage}{1\textwidth}
    \begin{subfigure}[t]{0.45\textwidth}
        \centering
        \includegraphics[width=\textwidth]{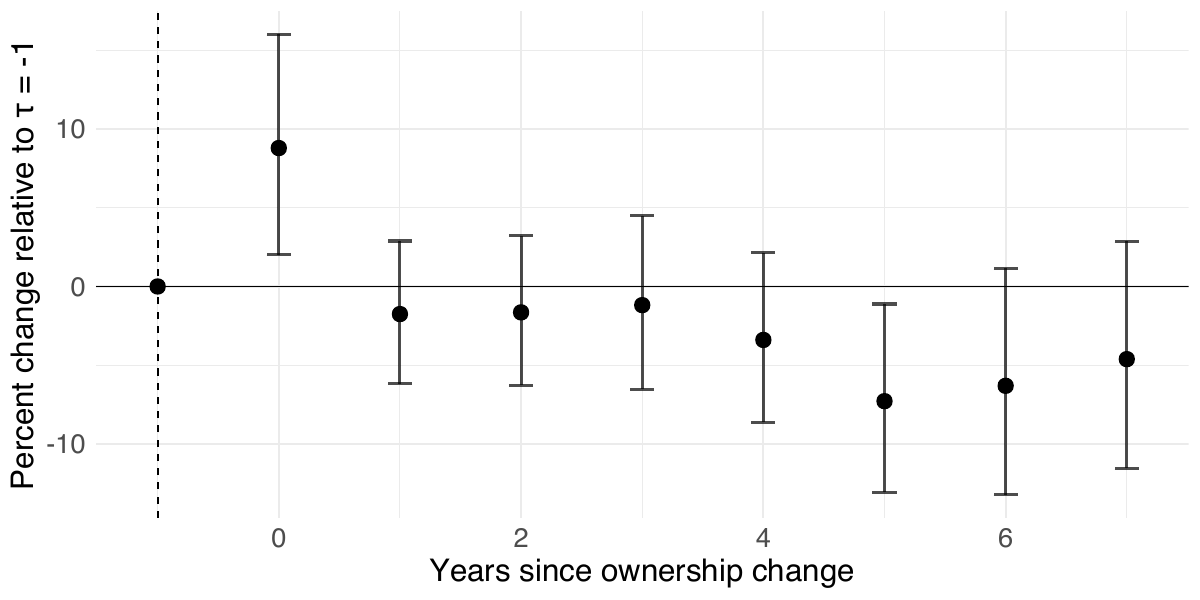}
        \caption{Ownership Change Event 1}
        \label{figure:OC1LogSalesEventStudy}
    \end{subfigure}%
    ~ 
    \begin{subfigure}[t]{0.45\textwidth}
        \centering
        \includegraphics[width=\textwidth]{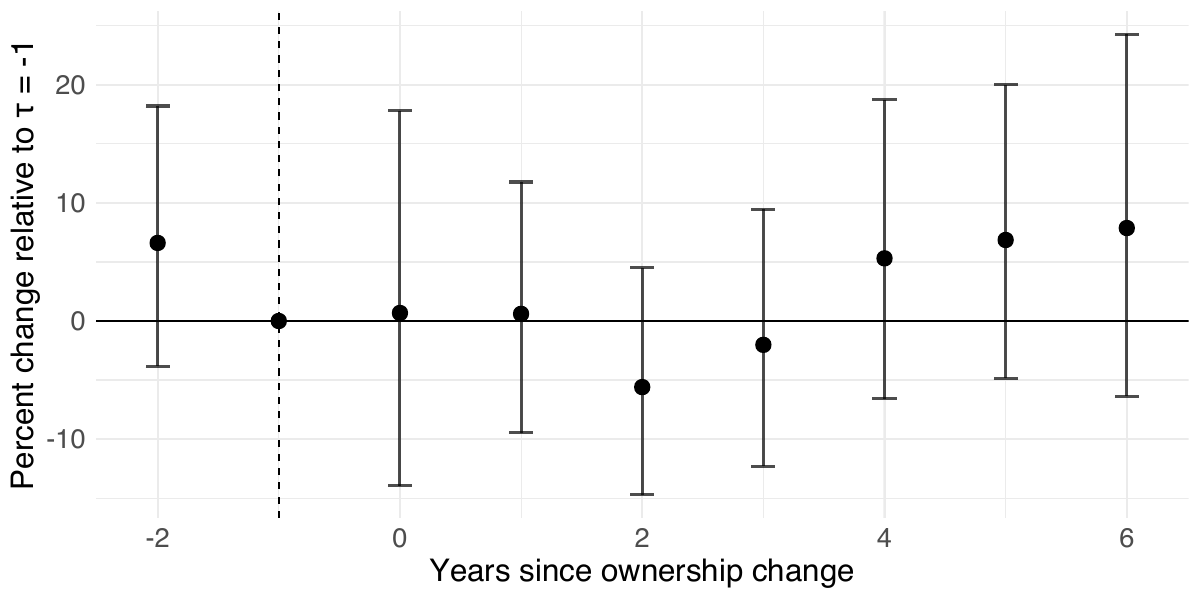}
        \caption{Ownership Change Event 2}
        \label{figure:OC2LogSalesEventStudy}
    \end{subfigure}
    ~  
    \begin{subfigure}[t]{0.45\textwidth}
        \centering
        \includegraphics[width=\textwidth]{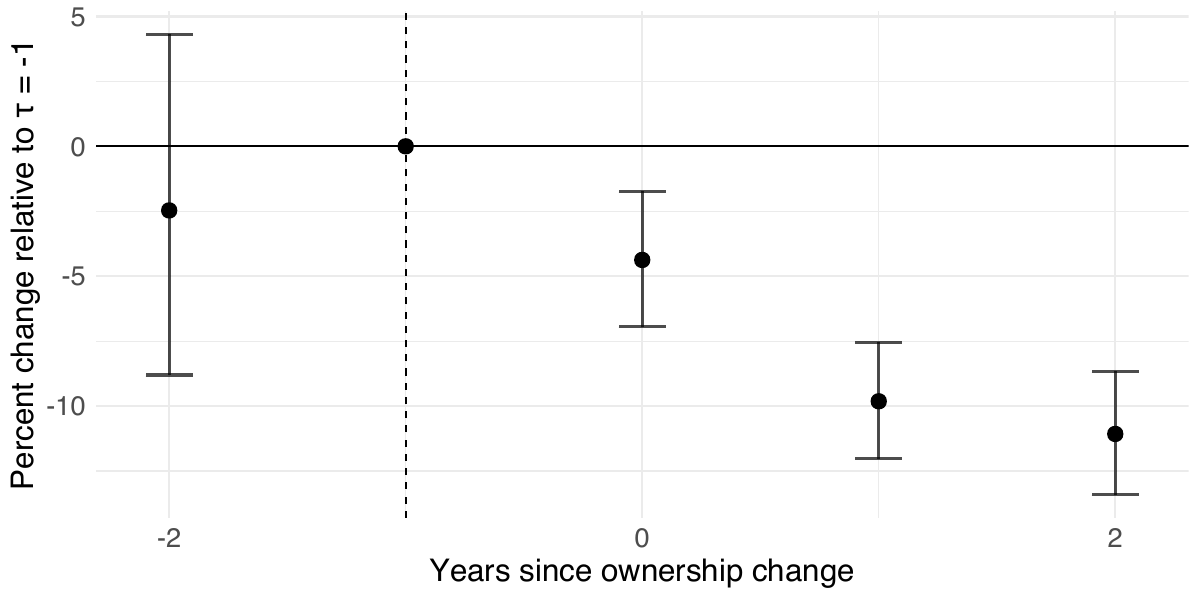}
        \caption{Ownership Change Event 3}
        \label{figure:OC3LogSalesEventStudy}
    \end{subfigure}
    \caption{Event Study Estimates of the Effect of Ownership Change on Store Sales}
    \label{figure:the.impact.of.OC.on.store.sales}
    \begin{flushleft}
        \footnotesize
        \textit{Notes:} Calculations are based on analysis of confidential investigation data from the FTC. The figures plot the event-study estimates of the effects of non-divestiture ownership change on log store sales for three anonymous non-divestiture acquisition events. To protect confidentiality, the identities of the acquisitions are not disclosed. The reported percentage changes are obtained by transforming log changes using $100 \times (e^{\hat{\beta}}-1)$.
    \end{flushleft}
    \end{minipage}
\end{figure}

\figref{figure:the.impact.of.OC.on.store.sales} presents event-study estimates associated with the effect of ownership change on store sales, and \tabref{table:ownership_change} columns (1)--(3) report their post-change averages relative to the year preceding the transaction. For two of the three events, we observe small, statistically insignificant declines in sales of approximately 2 percent in the years following the ownership change. For the third event, sales decline by roughly 8 percent, although only two years of post-change data are available. In all cases, these effects are substantially smaller than the sales declines observed following divestitures reported in \tabref{table:did_sales}.

\begin{table}[htbp]
\small
\centering
\caption{Effect of Ownership Change on Store Sales and Profit Margins}
\label{table:ownership_change}

\centering
\begin{threeparttable}
\begin{tabular}{lccccc}
\toprule
 & \multicolumn{3}{c}{Sales (\%)}  & \multicolumn{2}{c}{Profit Margin (p.p.)} \\ \cmidrule(lr){2-4} \cmidrule(lr){5-6}
 & Event 1 & Event 2 & Event 3 & Event 1 & Event 2 \\   
 & (1) & (2) & (3) & (4) & (5)  \\  
   \midrule
   Ownership Change & -2.3 & -1.8 & -8.4$^{***}$ & -0.89$^{***}$ & 1.24$^{*}$ \\   
   & (2.8) & (4.9) & (1.0)  & (0.30) & (0.70) \\   
   \midrule 
   $R^2$ & 0.895 & 0.931 & 0.990 & 0.819 & 0.736 \\  
\bottomrule
\end{tabular}
    \begin{tablenotes}
        \footnotesize
        \item \emph{Notes:} *** $p<0.01$, ** $p<0.05$, * $p < 0.10$. The dependent variable is either the log of sales excluding fuel and pharmacy, or profit margin defined as operating income divided by sales. Control stores are located in the same state as treated stores but in different three-digit ZIP codes. Sales coefficients are reported as percent changes, computed as $100(\exp(\hat\beta)-1)$; corresponding standard errors are obtained via the delta method. Profit margin coefficients are reported in percentage points.
    \end{tablenotes}
    \end{threeparttable}
\end{table}

\begin{figure}[htbp]
    \centering
    \begin{minipage}{1\textwidth}
    \begin{subfigure}[t]{0.45\textwidth}
        \centering
        \includegraphics[width=\textwidth]{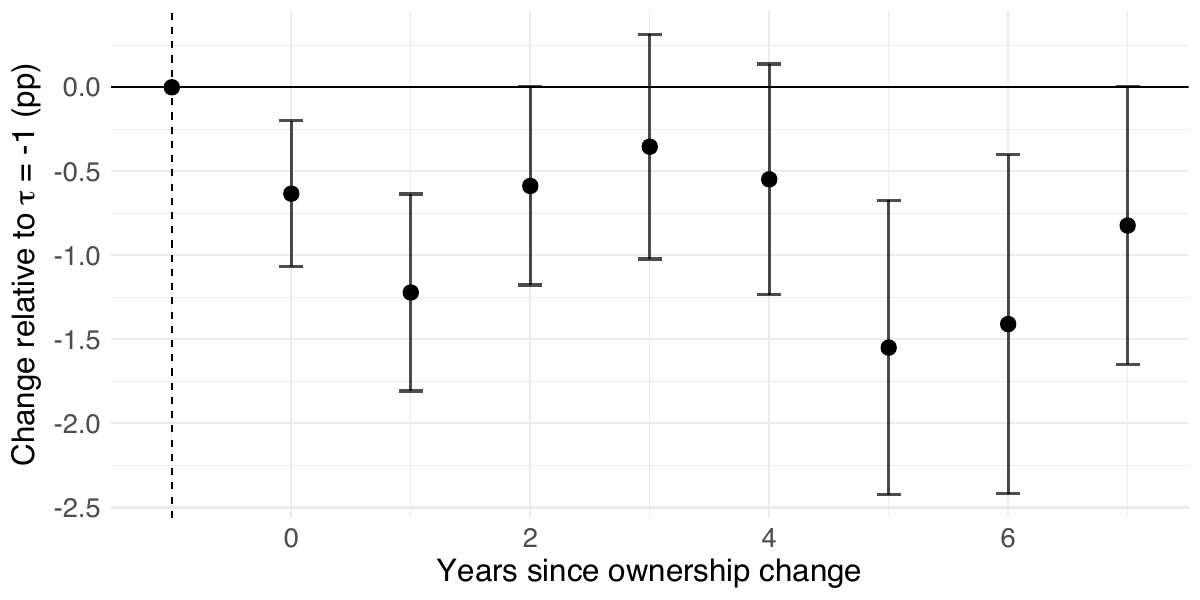}
        \caption{Event 1}
        \label{figure:OC1NMEventStudy}
    \end{subfigure}%
    \hfill
    \begin{subfigure}[t]{0.45\textwidth}
        \centering
        \includegraphics[width=\textwidth]{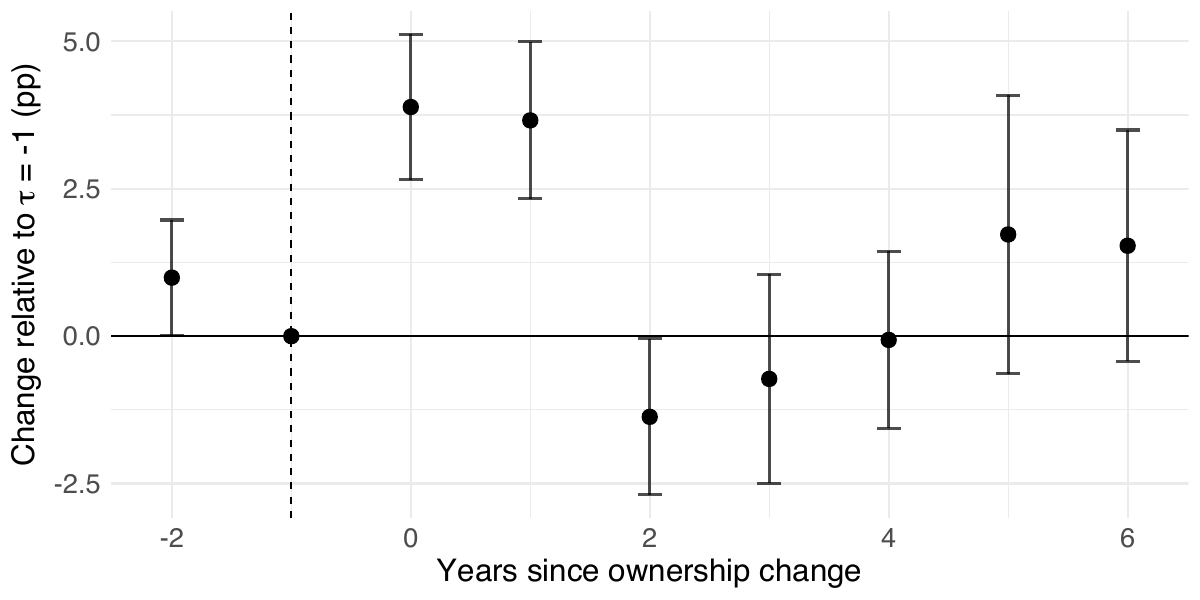}
        \caption{Event 2}
        \label{figure:OC2NMEventStudy}
    \end{subfigure}
      \caption{Event Study Estimates of the Effect of Ownership Change on Profit Margins}
    \label{figure:the.impact.of.OC.on.net.margin}        
    \begin{flushleft}
        \footnotesize
        \textit{Notes:} Calculations are based on analysis of confidential investigation data from the FTC. The figures plot the event-study estimates of the effects of non-divestiture ownership change on store profit margins for two anonymous non-divestiture acquisition events. To protect confidentiality, the identities of the acquisitions are not disclosed.
    \end{flushleft}
    \end{minipage}
    
\end{figure}

We next examine profit margins for two of the events.\footnote{Margin data for the third ownership-change event are unavailable.} \figref{figure:the.impact.of.OC.on.net.margin} presents the event-study estimates and \tabref{table:ownership_change} columns (4)--(5) report their post-change averages. For the first ownership change, margins decline by 0.9 percentage points; for the second, margins increase by 1.2 percentage points. Overall, we find little evidence that ownership turnover alone leads to sustained deterioration in sales or profitability. These results suggest that the large performance declines observed following divestitures cannot be attributed solely to ownership change.

\paragraph{Transition Costs}

A further component of transition dynamics involves the direct costs associated with rebranding and renovation following divestiture. These include expenditures on store remodeling, signage replacement, equipment upgrades, and integration into the buyer’s operating systems. Such costs may generate short-run disruptions and financial pressure that affect post-divestiture performance.

\tabref{table:distribution.of.remodeling.cost.share} reports remodeling expenditures as a share of average annual store sales for two divestiture transactions (Divestitures 1 and 2) in which many stores were re-bannered. For Divestiture 1, the mean remodeling cost equals 8.0 percent of annual revenue, while for Divestiture 2, the mean is 4.5 percent; median values are similar. Although costs vary across stores—with some exceeding 25 percent of annual revenue—all stores for Divestiture 2 and the majority of stores for Divestiture 1 incur remodeling costs below 10 percent of annual revenue.

\begin{table}[htbp]
\small
    \centering
    \caption{Divestiture Remodeling Cost Share of Sales}
    \label{table:distribution.of.remodeling.cost.share}
    \begin{threeparttable}
    \begin{tabular}{lcccccc} \toprule
    & Mean & Min & Q1 & Median & Q3 & Max \\ \midrule
    Divestiture 1  & 8.0\% & 0.0\% & 0.0\% & 8.0\% & 13.5\% & 28.7\% \\ 
    Divestiture 2 & 4.5\% & 3.6\% & 3.9\% & 4.2\% & 4.3\% & 8.1\% \\ 

    \bottomrule
    \end{tabular}
    \begin{tablenotes}
        \footnotesize
        \item \emph{Notes:} The table reports the distribution of the remodeling cost share of store sales in two divestiture transactions.
    \end{tablenotes}
    \end{threeparttable}
\end{table}

Because these expenditures are one-time and modest relative to annual sales, they are unlikely to explain the large and persistent sales declines documented earlier. While remodeling may contribute to short-run adjustment frictions, transition costs alone cannot account for the observed underperformance of divested stores.

\paragraph{Buyer Expectations}

Transition costs may be amplified if buyers fail to anticipate short-run performance declines following ownership transfer. Divestiture transactions often occur under compressed timelines, potentially limiting detailed due diligence and operational planning. If buyers project stable or improving sales paths, they may underinvest in working capital, staffing, or integration infrastructure during the transition period, thereby exacerbating short-run underperformance.

We examine internal projection data from one divestiture buyer. At acquisition, the buyer employed a simplified forecasting framework in which each acquired store was assigned either flat sales or 2 percent cumulative growth over four years. This approach generated nearly uniform projections across locations and implicitly assumed minimal transitional disruption.

\tabref{table:expected.vs.actual.sales} compares projected and realized sales two years after acquisition, based on a retrospective internal review conducted by the buyer. Aggregate realized sales were 8.5 percent below initial projections, and the median store underperformed its forecast by 5.1 percent. These systematic forecast errors indicate that the buyer did not fully anticipate the magnitude of transitional disruption.

\begin{table}[htbp]
\small
    \centering
    \caption{A Divestiture Buyer's Expected vs. Actual Sales After Two Years}
    \label{table:expected.vs.actual.sales}
    \begin{threeparttable}
    \begin{tabular}{lccccccc} \toprule
    & All & \multicolumn{6}{c}{Store-level} \\ \cmidrule(lr){2-2} \cmidrule(lr){3-8}
    & Total & Mean &  Min & Q1 & Median & Q3 & Max \\ \midrule
        Expectation Gap & $-8.5\%$ & $-2.7\%$ & $-32.2\%$ & $-16.4\%$ & $-5.1\%$ & $14.9\%$ & $26.5\%$ \\ \bottomrule
    \end{tabular}    
    \begin{tablenotes}
        \footnotesize
        \item \emph{Notes:} We define expectation gap as the actual store sales relative to predicted store sales. Column ``Total'' reports the total actual sales relative to the total predicted sales of the divested stores acquired by the divestiture buyer. We suppress the sample size to mask firm identity.
    \end{tablenotes}
    \end{threeparttable}
\end{table}

\begin{figure}[htbp]
    \centering
    \begin{minipage}{1\textwidth}
    \includegraphics[width=1\textwidth]{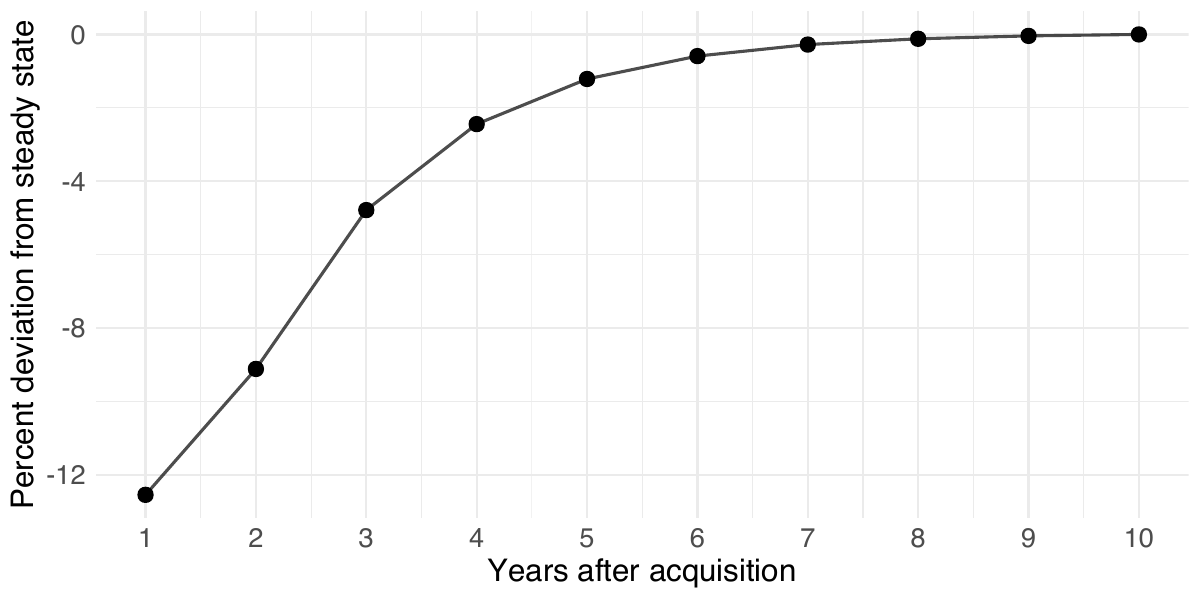}
    \caption{An Experienced Buyer's Post-Acquisition Sales Projections}
    \label{figure:an.experienced.firms.sales.projections}
    \begin{flushleft}
        \footnotesize
        \textit{Notes:} Calculations are based on analysis of confidential investigation data from the FTC. The figure plots the percentage deviation from the steady-state (final-period) value of average projected sales across acquired stores. The sales projections are generated by a prediction model of a large, experienced retail chain following an anonymous acquisition event. To preserve confidentiality, the identities of the firm and the transaction are not disclosed.
    \end{flushleft}
    \end{minipage}
\end{figure}

For comparison, \figref{figure:an.experienced.firms.sales.projections} presents projection paths from a larger and more experienced firm in a separate acquisition. That firm explicitly incorporated an initial sales shortfall of more than 12 percent relative to steady state and projected gradual convergence over a 7–10 year horizon. This contrast suggests that more experienced acquirers may internalize transition losses in their planning and capital allocation decisions.

Taken together, these patterns suggest that expectation errors may exacerbate short-run adjustment problems following divestiture. When buyers underestimate transitional disruption, the resulting gap between expected and realized performance may intensify adjustment frictions and delay recovery. However, in light of the modest magnitude of direct transition costs and the limited effects of ownership change alone, these dynamics are unlikely to fully explain the persistent underperformance documented earlier.

%% file: Conclusion.tex
\section{Conclusion \label{section:conclusion}}

This paper provides systematic evidence on the performance of divestiture remedies in supermarket mergers. Using establishment-level data, we show that divested stores are significantly more likely to exit and experience substantial declines in employment, sales, and profitability relative to comparable control stores. These patterns suggest that many stores do not remain viable competitors after divestiture.

We also examine the mechanisms underlying these outcomes. Both asset selection and buyer capability play important roles: divested stores tend to be weaker prior to transfer, and outcomes vary substantially with the characteristics of the acquiring firm. By contrast, transition frictions appear limited, suggesting that persistent underperformance reflects structural features of the remedy rather than temporary adjustment costs.

An important question for future research is whether these findings generalize beyond the supermarket industry. Grocery retail has several distinctive features, including reliance on distribution networks, the importance of private-label products, and substantial differentiation across price, quality, and store format, that may make it difficult for divestiture buyers to replicate the capabilities of incumbent chains. Evidence from other industries with different asset characteristics and buyer pools would help assess the generality of these findings. %

%% file: App.controls.tex
\section{Alternative Control Groups} \label{app:controls}

\begin{figure}[htbp!]
    \centering
    \begin{minipage}{1\linewidth}
    \includegraphics[width=1\linewidth]{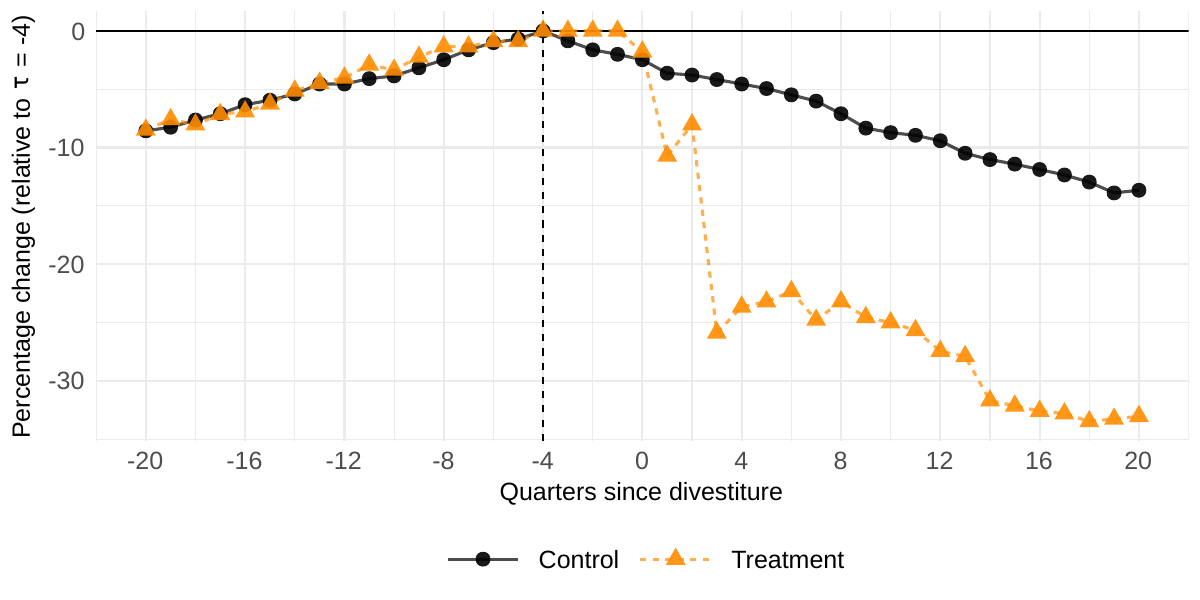}
    \caption{Change in Number of Stores Relative to Pre-Divestiture Baseline ($\tau = -4$)} \label{fig:survival}
    \begin{flushleft}
        \footnotesize
        \textit{Notes:} Calculations are based on data from the BLS QCEW LDB. The figure shows the percentage change in the number of active stores for the treated and control groups relative to four quarters preceding the divestiture date.
    \end{flushleft}
    \end{minipage}
\end{figure}

We begin by plotting the change in the number of stores in the treated and control groups over time (\figref{fig:survival}). For both groups, the series shows the percentage deviation from the baseline four quarters prior to divestiture. The two groups exhibit similar growth in the number of stores prior to divestiture. Following divestiture, however, the number of stores declines sharply for the treated group, falling by 24 percent in the first year compared with only a 4.5 percent decline for control stores. Over the five years following divestiture, 14 percent of control stores exit compared with 33 percent of divested stores.

Next, we examine the sensitivity of the estimates in \tabref{table:did_bls} to the choice of control group. We consider four alternative specifications that modify the matching restrictions used in the main analysis.

First, \tabref{table:did_age_window_20q} expands the allowable age difference between treated and control stores to 20 quarters. Second, \tabref{table:did_employment_window_40pct} expands the allowable employment difference between treated and control stores to 40 percent. Third, \tabref{table:did_n5} uses the five closest stores rather than the ten used in the main specification. Finally, \tabref{table:did_n15} uses the fifteen closest stores.

The estimates across these specifications are similar to the baseline results. This robustness likely reflects the relatively stable employment and payroll patterns of stores belonging to established chains, so modest changes in matching criteria have little effect on the estimated treatment effects. Overall, these results indicate that the baseline estimates are not sensitive to the specific control group construction.

\begin{table}[htbp]
\small
\centering
\caption{Horizon-Specific Effects of Divestiture: Age Window 20 Quarters}
\label{table:did_age_window_20q}
\begin{threeparttable}
\begin{tabular}{lccccc}
\toprule
 & Survival (pp) & Employment (\%) & DHS Growth Rate (pp) & Payroll (\%) & Avg Earnings (\%) \\
Post Horizon & (1) & (2) & (3) & (4) & (5) \\ \midrule
1 Year & -11.9$^{***}$ & -22.5$^{***}$ & -47.9$^{***}$ & -21.7$^{***}$ & 1.1 \\
 & (2.4) & (2.7) & (4.6) & (2.6) & (2.2) \\
2 Years & -14.3$^{***}$ & -19.0$^{***}$ & -50.1$^{***}$ & -21.9$^{***}$ & -3.6$^{*}$ \\
 & (2.5) & (2.4) & (4.3) & (2.5) & (2.0) \\
5 Years & -16.7$^{***}$ & -15.8$^{***}$ & -53.2$^{***}$ & -21.2$^{***}$ & -6.4$^{***}$ \\
 & (2.5) & (2.8) & (4.6) & (2.9) & (2.1) \\
\bottomrule
\end{tabular}
    \begin{tablenotes}
        \footnotesize
        \item \emph{Notes:} *** $p<0.01$, ** $p<0.05$, * $p < 0.10$. The table reports the horizon-specific average treatment effects calculated via equation \eqref{eq:lin_com}. The numbers are calculated from the BLS QCEW LDB. Control stores are defined as described in \secref{sub:control}, except that the age difference between divested and control stores can be up to 20 quarters. Survival and DHS growth rate are reported in percentage points. Employment, payroll, and average earnings are reported as percent changes, computed as $100(\exp(\hat\beta)-1)$; standard errors are obtained via the delta method.
    \end{tablenotes}
\end{threeparttable}
\end{table}

\begin{table}[htbp]
\small
\centering
\caption{Horizon-Specific Effects of Divestiture: Employment Window 40 Percent}
\label{table:did_employment_window_40pct}
\begin{threeparttable}
\begin{tabular}{lccccc}
\toprule
 & Survival (pp) & Employment (\%) & DHS Growth Rate (pp) & Payroll (\%) & Avg Earnings (\%) \\
 Post Horizon & (1) & (2) & (3) & (4) & (5) \\ \midrule
1 Year & -12.8$^{***}$ & -21.8$^{***}$ & -47.1$^{***}$ & -21.1$^{***}$ & 0.9 \\
 & (2.2) & (2.7) & (4.8) & (2.5) & (2.1) \\
2 Years & -14.3$^{***}$ & -18.5$^{***}$ & -49.3$^{***}$ & -21.7$^{***}$ & -3.8$^{**}$ \\
 & (2.4) & (2.2) & (4.5) & (2.4) & (1.9) \\
5 Years & -15.9$^{***}$ & -15.8$^{***}$ & -52.1$^{***}$ & -21.3$^{***}$ & -6.6$^{***}$ \\
 & (2.4) & (2.7) & (4.9) & (2.8) & (2.0) \\
\bottomrule
\end{tabular}
    \begin{tablenotes}
        \footnotesize
        \item \emph{Notes:} *** $p<0.01$, ** $p<0.05$, * $p < 0.10$. The table reports the horizon-specific average treatment effects calculated via equation \eqref{eq:lin_com}. The numbers are calculated from the BLS QCEW LDB. Control stores are defined as described in \secref{sub:control}, except that the difference in employment between the divested and control stores can be up to 40 percent. Survival and DHS growth rate are reported in percentage points. Employment, payroll, and average earnings are reported as percent changes, computed as $100(\exp(\hat\beta)-1)$; standard errors are obtained via the delta method.
    \end{tablenotes}
\end{threeparttable}
\end{table}

\clearpage 

\begin{table}[htbp]
\small
\centering
\caption{Horizon-Specific Average Effects of Divestiture (5 Control Stores)} \label{table:did_n5}
\begin{threeparttable}
\begin{tabular}{lccccc}
\toprule
 & Survival (pp) & Employment (\%) & DHS Growth Rate (pp) & Payroll (\%) & Avg Earnings (\%) \\
Post Horizon & (1) & (2) & (3) & (4) & (5) \\ \midrule
1 Year & -13.0$^{***}$ & -21.6$^{***}$ & -47.5$^{***}$ & -21.3$^{***}$ & 0.4 \\
 & (2.2) & (2.7) & (4.7) & (2.6) & (2.2) \\
2 Years & -15.2$^{***}$ & -18.0$^{***}$ & -49.6$^{***}$ & -21.5$^{***}$ & -4.4$^{**}$ \\
 & (2.3) & (2.3) & (4.3) & (2.5) & (1.9) \\
5 Years & -17.6$^{***}$ & -14.9$^{***}$ & -52.6$^{***}$ & -21.0$^{***}$ & -7.1$^{***}$ \\
 & (2.3) & (3.0) & (4.8) & (3.0) & (1.9) \\
\bottomrule
\end{tabular}
    \begin{tablenotes}
        \footnotesize
        \item \emph{Notes:} *** $p<0.01$, ** $p<0.05$, * $p < 0.10$. The table reports the horizon-specific average treatment effects calculated via equation \eqref{eq:lin_com}. All estimates are based on data from the BLS QCEW LDB. Control stores are defined as described in \secref{sub:control}, except that we select the five closest stores. Survival and DHS growth rate are reported in percentage points. Employment, payroll, and average earnings are reported as percent changes, computed as $100(\exp(\hat\beta)-1)$; standard errors are obtained via the delta method.
    \end{tablenotes}
\end{threeparttable}
\end{table}

\begin{table}[htbp]
\small
\centering
\caption{Horizon-Specific Average Effects of Divestiture (15 Control Stores)} \label{table:did_n15}
\begin{threeparttable}
\begin{tabular}{lccccc}
\toprule
 & Survival (pp) & Employment (\%) & DHS Growth Rate (pp) & Payroll (\%) & Avg Earnings (\%) \\
Post Horizon & (1) & (2) & (3) & (4) & (5) \\ \midrule
1 Year & -12.3$^{***}$ & -22.1$^{***}$ & -46.9$^{***}$ & -21.0$^{***}$ & 1.4 \\
 & (2.3) & (2.8) & (4.6) & (2.5) & (2.1) \\
2 Years & -14.4$^{***}$ & -18.6$^{***}$ & -49.2$^{***}$ & -21.5$^{***}$ & -3.5$^{*}$ \\
 & (2.4) & (2.4) & (4.3) & (2.4) & (1.8) \\
5 Years & -16.4$^{***}$ & -14.7$^{***}$ & -51.7$^{***}$ & -20.3$^{***}$ & -6.6$^{***}$ \\
 & (2.3) & (2.8) & (4.4) & (2.8) & (1.9) \\
\bottomrule
\end{tabular}
    \begin{tablenotes}
        \footnotesize
        \item \emph{Notes:} *** $p<0.01$, ** $p<0.05$, * $p < 0.10$. The table reports the horizon-specific average treatment effects calculated via equation \eqref{eq:lin_com}. All estimates are based on data from the BLS QCEW LDB. Control stores are defined as described in \secref{sub:control}, except that we select the 15 closest stores. Survival and DHS growth rate are reported in percentage points. Employment, payroll, and average earnings are reported as percent changes, computed as $100(\exp(\hat\beta)-1)$; corresponding standard errors are obtained via the delta method.
    \end{tablenotes}
\end{threeparttable}
\end{table}